\documentstyle[12pt]{article}
\newcommand{\be}{\begin{equation}}
\newcommand{\ee}{\end{equation}}
\newcommand{\bea}{\begin{eqnarray}}
\newcommand{\eea}{\end{eqnarray}}
\newcommand{\bn}{\begin{eqnarray*}}
\newcommand{\en}{\end{eqnarray*}}
\newcommand{\pdag}{{\phantom{\dagger}}}

\newcommand{\p}{\partial}
\newcommand{\s}{\sigma}

\newcommand{\up}{\uparrow}
\newcommand{\down}{\downarrow}
\newcommand{\la}{\langle}
\newcommand{\ra}{\rangle}
\newcommand{\rd}{\mbox{d}}
\newcommand{\ri}{\mbox{i}}

\newcommand{\var}{\varepsilon}

\begin {document}
\title{Critical properties of the double-frequency sine-Gordon model 
with applications}
\author{M. Fabrizio$^a$, A. O. Gogolin$^b$ and A. A. Nersesyan$^{c,d}$}
\vspace{2cm}
\maketitle

\begin{center}
$^a$International School for Advanced Studies and INFM, 
Via Beirut 4, 34014 Trieste, Italy\\ 
$^b$Department of Mathematics, Imperial College, 180 Queen's Gate,
London\\ 
SW7 2BZ, United Kingdom\\
$^c$ The Abdus Salam International Centre for Theoretical Physics,
P.O.Box 586, 34100, Trieste, Italy\\
$^d$The Andronikashvili Institute of Physics, Tamarashvili 6, 380077, 
Tbilisi, Georgia
\end{center}

\begin{abstract}

We study the properties of the double-frequency sine--Gordon model 
in the vicinity of the 
Ising quantum phase transition displayed by this model.  
Using a mapping onto a generalised 
lattice quantum Ashkin-Teller model, we obtain critical and
nearly-off-critical correlation functions of various operators.
We discuss applications of the double-sine-Gordon
model to one-dimensional physical systems, like 
spin chains in a staggered external field and interacting
electrons in a staggered potential.
\end{abstract}


\section{Introduction}

The problem of determining the asymptotic behaviour of a conformal field 
theory (CFT) under the action of a relevant operator is well studied and 
understood, also in view of its relation to the physics of many 
quantum one-dimensional (1D) and classical two-dimensional (2D) 
models. 
A less studied case concerns a CFT
perturbed by two relevant operators. 
Namely, consider a CFT subjected to such relevant 
perturbation, $g O_{\Delta_g}$, 
with scaling dimension $\Delta_g < 2$,
that turns
it into a fully massive quantum field
theory (QFT).
Then add another relevant perturbation,
$\lambda O_{\Delta_\lambda}$,
which, if acting alone, would also make
our QFT fully massive.
Although this is not as common situation as the previous one, 
we will show that it displays interesting features which 
may be realized in physical systems.

Without loss of generality, we
assume that $\Delta_\lambda<\Delta_g<2$,
i.e. that the second perturbation is more relevant
then the first one. Moreover, 
we shall require, for the time being, that the operator
product expansion (OPE) of these two
operators is closed in the sense that it
does not produce other relevant operators. The most general case 
will be discussed in a separate section. 
Then, the na\"{i}ve expectation, which prevailed until recently,
is that no qualitative changes occur in this case with respect to the 
standard situation of a CFT perturbed by a single relevant operator:
the low-energy behaviour of the theory
would be governed by the most relevant operator
and, at any rate, the theory would remain 
fully massive. 
This expectation is however not a general rule if the 
two operators {\sl exclude} each other, that is, if the 
field configurations which minimise one perturbation term 
do not minimise the other. 
In this case, the interplay between the two competing
relevant operators  
can produce a novel quantum phase transition between two massive QFT's 
through a critical (massless) point. It is intuitively clear that
such a scenario is only conceivable if the ratio of the
bare coupling constants, $|g/\lambda|$, is large enough. 

Recently Delfino and Mussardo (DM)\cite{DM} considered
the double-frequency sine-Gordon (DSG) model, which is a 
Gaussian model of a scalar field $\Phi$, 
perturbed by two relevant vertex operators with
the ratio of their scaling dimensions $\Delta_g/\Delta_{\lambda}=4$.
The Hamiltonian density of the DSG model reads:
\bea
{\cal H}_{DSG}[\Phi] &=& {\cal H}_0[\Phi] + {\cal U}[\Phi]\nonumber\\
 {\cal U}[\Phi] &=&
- g\cos\beta\Phi(x)-\lambda\sin[(\beta/2)\Phi(x)].
\label{HDSG}
\eea
where 
\be
{\cal H}_0[\Phi]=v_0\left[
(\p_x\phi_R)^2+(\p_x\phi_L)^2\right],
\label{HPhifree}
\ee
$\phi_{R,L}$ being the chiral components of the Bose field,
$\Phi=\phi_R+\phi_L$.
DM have shown that there exists a quantum critical 
line $\lambda=\lambda_c(g)$ where the DSG model displays
an Ising criticality with central charge $c=1/2$.

The purpose of this paper is to investigate in detail 
the critical properties of this 
transition and to discuss physical applications
to spin chains and other quantum one-dimensional (1D)
systems. While we shall concentrate on the DSG model in what follows,
it should be noted that the phenomenon is more general - similar quantum
phase transitions do happen in more complicated models;
those shall be addressed elsewhere \cite{else}.

Apart from the practical interest in physical realizations of the DSG model,
this problem is of relevance also from a pure 
theoretical point of view. Indeed, 
we have already mentioned that the Ising critical point separates 
two {\sl strong-coupling}, massive phases; hence, by definition,
its analytical description is outside the range of applicability 
of perturbation theory. In this paper we propose a nonperturbative 
scheme, which is essentially based on the mapping of the DSG model onto 
another equivalent model -- a generalised quantum Ashkin-Teller model
on a 1D lattice, where the Ising critical point becomes accessible.

The paper is organised as follows. In section 2 we briefly discuss 
the DSG model at the semi-classical level and outline our approach
to tackle the Ising criticality in the vicinity of the decoupling
point $\beta^2 = 4\pi$. 
In sections 3--5 we present 
a complete description of the critical properties of the DSG model
at the Ising transition.
In section 3 we introduce a quantum lattice
version of a generalised AT model which, apart from
the conventional marginal inter-chain coupling, also includes a
`magnetic filed' type of interaction, $h \s_1 \s_2$. We show that,
in the continuum limit, this model is equivalent to the DSG model
at $\beta^2 \sim 4\pi$. In section 4 we employ a new 
($\s$-$\tau$) representation of the deformed AT model 
which enables us
to correctly identify those degrees of freedom that become critical.
 In section 5
we use the results of the quantum-spin-chain mappings to determine
the physical properties of the DSG model close to the criticality.

In the next two
sections we discuss applications of the DSG
model to physical systems. In particular, we discuss Ising transitions
in a spontaneously dimerized S=1/2 antiferromagnetic chain in a staggered
magnetic field (section 6), and in the 1D Hubbard model with a staggered
potential (section 7). 
In section 8 we show that at $\beta^2 < 2\pi$ the second-order Ising transition
may transform to a first-order one. Section 9 contains discussion of the 
results and conclusions. 

The paper is supplied with four Appendices which provide some details 
concerning
bosonization of products of order and disorder operators for a system of two
identical Ising models, the relationship between the quantum AT model and
lattice fermions, a mean-field treatment of the $\s$-$\tau$ model,
and calculation of the correlation functions.

\section{The model and its quasi-classical analysis}

In this paper we will be dealing with
the particular version of the DSG model (\ref{HDSG}) 
corresponding to the case $g >0$.
 We will also assume that $\lambda > 0$, although its
sign is unimportant due to the obvious symmetry $\Phi \to -\Phi$. 

For $2\pi<  \beta^2 < 8\pi$ the Gaussian model is perturbed by
two relevant vertex operators closed under the OPE \cite{DM}.
The existence of the Ising phase transition
can be qualitatively understood via
a quasi-classical analysis: inspecting the profile
of the potential ${\cal U} (\Phi)$ as a function of the ratio $x=\lambda/4g$.
At $x = 0$ one has a 
periodic potential of a single-frequency sine-Gordon (SG) model.
For $x \neq 0$ the period of the potential is doubled in such a way that
in the region $0 < x < 1$ 
it can be viewed as a sequence of double-well potentials
with the local structure $A \Phi^2 + B \Phi^4$, where $A<0, B>0$.
Precisely at $x=1$ each local double-well 
potential transforms to a $\Phi^4$ one
$(A=0)$, and (in the Ginzburg--Landau sense) 
this is a signature of the Ising criticality
with the central charge $c=1/2$.
It should be stressed that the double-well structure of the potential
${\cal U}(\Phi)$ only occurs for $g > 0$. 
Hence the Ising critical point only exist  for
positive $g$. The case of $g < 0$ is qualitatively different:
here the $\lambda$-perturbation removes the degeneracy between neighbouring
minima of $\cos\beta\Phi$ and thus leads to soliton 
confinement (similarly to the analysis in Ref\cite{aff-conf}). 
The spectrum in this case always remains massive. 

Thus, for model (\ref{HDSG}) with $g > 0$,
a plausible scenario is that
the relevant perturbations 
naturally act on the two constituent 
Ising models of the starting Gaussian model,
leaving one of them massless on the
critical line $\lambda=\lambda_c(g)$.
(Quasi-classically $\lambda_c(g)=4g$. A better estimation which takes into
account quantum fluctuations yields a power law $\lambda_c(g) \sim g^{\nu}$,
where $\nu = \left(32\pi - \beta^2 \right)/\left(16\pi - \beta^2 \right)$).
Indeed,
DM have argued that the Ising transition is a
universal property of the DSG model (\ref{HDSG}) as long as $\beta^2 < 8\pi$.
This makes it possible to consider the vicinity of the point
$\beta^2 = 4\pi$ where the description of the transition greatly simplifies.
Namely,
it is well known\cite{Coleman,LE} that $\beta^2 = 4\pi$ is the decoupling
(or Luther-Emery) point of the ordinary SG model ($\lambda = 0$),
at which the latter is equivalent to a theory of free massive fermions.
The scaling dimension 1/4 of the $\lambda$-term in (\ref{HDSG}) indicates
that this point is special for the DSG model as well, suggesting 
an Ising model interpretation. In this paper we shall be working in 
the vicinity of the decoupling point. Rescaling the field $\Phi$,
the DSG model can be written in an equivalent form:
\bea
{\cal H}_{DSG}&=&v\left[
(\p_x\phi_R)^2+(\p_x\phi_L)^2\right]
-\gamma\p_x\phi_R(x)\p_x\phi_L(x)\nonumber\\&-&
g\cos\sqrt{4\pi}\Phi(x)-\lambda\sin\sqrt{\pi}\Phi(x),
\label{HDSGbis}
\eea
where $v=(v_0/2)(\beta/\sqrt{4\pi}+\sqrt{4\pi}/\beta)$
and $\gamma=(v_0/\sqrt{4\pi}\beta)(\beta^2-4\pi)$.
It was originally noted by DM that this form of
the DSG model can be related to a deformed 2D Ashkin--Teller (AT)
model, which can be viewed as the standard  AT model, i.e.
a model of two marginally coupled 2D Ising models
with order parameters $\s_1$ and $\s_2$, extended to
include a magnetic field type coupling, $\lambda \s_1 \s_2$.
DM then argued that, in the strong-coupling limit
($\lambda \rightarrow \infty$), the system effectively reduces
to a single Ising model which may become critical if the
temperature is properly tuned to its critical value.
This argument can be extended to finite values of $\lambda$,
provided that the two Ising copies are originally
in a disordered phase, and thus explains the existence of a critical line
in the $t$-$\lambda$ plane ($t = (T - T_c)/T_c > 0$)
along which the deformed AT model flows from the ultraviolet
$c = 1$ fixed point to the infrared $c = 1/2$ Ising fixed point.

We shall elaborate on the DM argument
by explicitly constructing the mappings between the
DSG model and various spin models, with the aim to
fully describe the Ising transition and, in particular,
calculate correlation functions
of the operators in the original DSG model.
The novel points of our analysis are as follows.
First, we concentrate on quantum lattice spin chain
models
rather than on their classical counterparts
(in the transfer matrix sense), as this formulation
profits from the use of the powerful apparatus
of various spin operator transformations like
the duality transformation. 
Secondly, we identify the correct degrees of freedom in
the deformed quantum AT model 
that become critical. 
Since the original Ising models enter symmetrically,
this is a nontrivial step 
which is accomplished via a `change of basis' 
transformation. Let us denote the critical degrees of freedom by
$\sigma$ and the remaining gaped degrees of freedom
by $\tau$ (see the main text for precise definitions).
The strategy to calculate the correlation functions
is then to express the DSG-operators in terms of 
the lattice $\s$ and $\tau$ operators. At the Ising transition
$\s$ and $\tau$ operators asymptotically decouple, 
with the $\s$-operators being critical and (some of) 
the $\tau$ operators acquiring finite average values. 
This allows us to trace the relation
between the original DSG-operators and those 
from the operator content of the underlying
critical Ising model, which ultimately accomplishes a
complete description of the critical properties at the
Ising transition.

\section{Relation between DSG model and deformed quantum Ashkin-Teller
model}

\subsection{Quantum Ising spin chain}

We start with recollecting some basic facts about the
quantum Ising (QI) spin chain. The Hamiltonian of the QI chain
describes a 1D Ising model in a transverse magnetic
field \cite{Pf}:
\be
H_{QI}[\sigma]=-\sum_n\left(J\s^z_n\s^z_{n+1}+\Delta \s^x_n
\right),
\label{QI}
\ee
where $\s^{\alpha}_n$ are the Pauli matrices associated with the lattice sites
$\{n \}$. The Hamiltonian $H_{QI}$ defines the transfer
matrix of the classical 2D Ising model 
\cite{LSM}.

An important tool in studying 1D spin lattice models
is the Kramers-Wannier duality transformation that
we shall make extensive use of in the sequel. Consider
a dual lattice consisting of sites 
$\{n +1/2\}$, 
defined as the centres of the links $< n,n+1>$
of the original lattice,  and assign spin operators
$\mu_{n+1/2}$ to the dual lattice sites.
The duality transformation then
relates the dual spins to the original ones as
follows
\be
\mu^z_{n+1/2}=\prod\limits_{j=1}^n\s^x_j,\;\;\;
\mu^x_{n+1/2}=\s^z_n\s^z_{n+1}, 
\label{dual}
\ee
the inverse transformation being
\be
\s^z_{n}=\prod\limits_{j=0}^{n-1}\mu^x_{j+1/2},\;\;\;
\s^x_{n}=\mu^z_{n-1/2}\mu^z_{n+1/2}.
\label{dualinv}
\ee 
In Hamiltonian (\ref{QI}), the parameters $J$ and
$\Delta$ are interchanged under the duality transformation, so
that $J=\Delta$ is the self-duality point where the
model displays an Ising criticality.

The operators $\s^z_n$ and $\mu^z_{m+1/2}$, conventionally referred
to as order and disorder operators, 
are mutually nonlocal and
play the role
of `string operators' with respect to each other. In
particular,
they commute for $m<n$ but anticommute
otherwise.
These commutation properties make the duality
construction
a convenient starting point for introducing lattice
fermions. It is immediate to check that two objects
\bea
\eta_n &=& \s^z _n \mu^z _{n-1/2} = \mu^z _{n-1/2}\s^z _n, \label{d1}\\
\zeta_n &=& \ri \s^z _n \mu^z _{n+1/2} = - \ri \mu^z _{n+1/2}\s^z _n,
\label{d2}
\eea
satisfy the 
anticommutation relations for the
real (Majorana) fermions on the lattice, with the
normalisation
$\eta^2_n=\zeta^2_n=1$. Relations (\ref{d1}) and (\ref{d2}) 
are nothing but the inverse
of the Jordan--Wigner transformation, with
the `direct' transformation being of the form
\be
\s^x_n=i\zeta_n\eta_n,\;\;\;
\s^z_n=\eta_n\prod\limits_{j=1}^{n-1}\left(i\zeta_j\eta_j\right).
\label{JW}
\ee

In terms of the Majorana fermions the QI Hamiltonian
becomes
\be
H_{QI}= \ri \sum\limits_n\left[J\zeta_n(\eta_{n+1}-\eta_n)-
(\Delta - J)\zeta_n\eta_n\right].
\label{QIM}
\ee
The next step is to take a continuum limit. 
To this end one introduces a lattice 
spacing $a_0$, treats $x=na_0$ as a continuum
variable, and replaces $\eta_n$ and $\zeta_n$ by slowly varying Majorana fields,
$\eta (x)$ and $\zeta (x)$:
\[
\eta_n \to \sqrt{2a_0} \eta(x),\;\;\;
\zeta_n \to \sqrt{2a_0} \zeta(x).
\]
Notice that the factor $\sqrt{2}$ ensures the correct
continuum anticommutation relations:
$\{\eta(x),\eta(y)\}= \{\zeta(x),\zeta(y) \}
= \delta(x-y)$. The Hamiltonian density of (\ref{QIM}) then is
\[
{\cal H}_{QI} =   \ri v \zeta\p_x\eta- \ri m\zeta\eta,
\]
with $v=2Ja_0$ and $m=2(\Delta-J)$.
Performing a chiral rotation of the Majorana spinor
\be
\xi_R = \frac{- \eta + \zeta}{\sqrt{2}}, ~~~
\xi_L = \frac{\eta + \zeta}{\sqrt{2}}, \label{chiral-rot1}
\ee 
or, inversely,
\be
\eta = \frac{- \xi_R + \xi_L}{\sqrt{2}}, ~~~
\zeta = \frac{\xi_R + \xi_L}{\sqrt{2}},\label{chiral-rot2}
\ee
transforms this Hamiltonian to a standard form:
\be
{\cal
H}^{(m)}_M=\frac{\ri v}{2}(-\xi_R\p_x\xi_R+\xi_L\p_x\xi_L)
-\ri m\xi_R\xi_L.
\label{Mfree}
\ee
The relation between the QI model and the massive Majorana
QFT
is therefore summarised as follows
\be
\lim\limits_{a_0\to 0}H_{QI}=\int \rd x~ {\cal H}^{(m)}_M (x)
\label{QIvsM}
\ee

\subsection{Bosonization of the deformed quantum Ashkin--Teller model}

The standard quantum Ashkin-Teller (QAT) model is defined as a model
of two identical QI spin chains, described by Hamiltonians
$H_{QI}[\s_1]$ and $H_{QI}[\s_2]$ [see Eq.(\ref{QI})],
which are coupled via a self-dual inter-chain interaction:
\bea
H_{QAT}&=&H_{QI}[\s_1]+H_{QI}[\s_2]+H'_{AT}[\s_1,\s_2],\label{QAT}\\
H'_{AT}[\s_1,\s_2]&=&K\sum\limits_n
\left(\s^z_{1,n}\s^z_{1,n+1}\s^z_{2,n}\s^z_{2,n+1}+
\s^x_{1,n}\s^x_{2,n}\right).
\label{QATint}
\eea
We will be interested in a deformed version of this model
which, apart from $H'_{AT}$, includes a  
`magnetic field' type of coupling between the chains:
\be
H_{DQAT} = H_{QAT} - h\sum\limits_n \s^z_{1,n}\s^z_{2,n}
\label{DQAT}
\ee

We wish to establish a relationship between the 
deformed quantum Ashkin--Teller (DQAT) model (\ref{DQAT}) 
and the DSG model. These two models can be mapped onto each other
using the Zuber-Itsykson trick\cite{ZI}. 
The idea is to associate the two QI chains with two copies 
of Majorana fermions,
\bea
\left( \s_1 , \mu_1 \right) &\Rightarrow& \left( \eta^1, \zeta^1 \right)
\Rightarrow \left( \xi^1 _R , \xi^1 _L \right), \nonumber\\
\left( \s_2 , \mu_2 \right) &\Rightarrow& \left( \eta^2, \zeta^2 \right)
\Rightarrow \left( \xi^2 _R , \xi^2 _L \right), \nonumber 
\eea
combine $\xi^1$ and $\xi^2$
into a single Dirac field and then bosonize the latter using the standard
rules of Abelian bosonization.

First we notice that, in the case of two QI 
spin chains, the Jordan--Wigner transformation (\ref{d1})--(\ref{JW})
should be slightly modified. This follows from
the requirement that the spin operators belonging to different chains
should commute. While this is automatically true for the disorder operators
$\mu^{\alpha}_{a, n+1/2}~(\alpha=x,y,z;~a=1,2)$ because of their
bosonic character, to ensure commutation between 
the order parameters
$\s^{z}_{1 n}$ and $\s^{z}_{2 n}$
one has to introduce two anticommuting (Klein) factors
\be
\{ \kappa_1 , \kappa_2 \} = 0, ~~~ \kappa^2 _1 = \kappa^2 _2 = 1,
\label{Klein-algebra}
\ee
and replace (\ref{d1}), (\ref{d2}) by
\bea
\eta_{1,n}&=&\kappa_1\s^z_{1,n}\mu^z_{1,n-1/2},\;\;\;
\zeta_{1,n}=i\kappa_1\s^z_{1,n}\mu^z_{1,n+1/2},\nonumber\\
\eta_{2,n}&=&\kappa_2\s^z_{2,n}\mu^z_{2,n-1/2},\;\;\;
\zeta_{2,n}=i\kappa_2\s^z_{2,n}\mu^z_{2,n+1/2}.
\label{Majbis}
\eea
With the spin variables $\s_{1,2}$ and $\mu_{1,2}$ subject to the duality 
relations
(\ref{dual}), (\ref{dualinv}), 
definitions (\ref{Majbis}) ensure the correct statistics for
the Majorana fermions $\eta_{a,n}$ and $\zeta_{a,n}$.

In terms of the lattice Majorana fermions, the standard QAT 
Hamiltonian becomes
\bea
H_{QAT}&=& \ri \sum\limits_{a=1,2}\sum\limits_n
\left[J\zeta_{a,n}(\eta_{a,n+1}-\eta_{a,n})-
(\Delta -J)\zeta_{a,n}\eta_{a,n}\right]\nonumber\\
&+&
K\sum\limits_n\zeta_{1,n}\zeta_{2,n}\left[
\eta_{1,n+1}\eta_{2,n+1}+\eta_{1,n}\eta_{2,n}\right],
\label{QAT-M}
\eea
and admits a straightforward passage to
the continuum limit. The corresponding Hamiltonian density 
reduces to a theory of two interacting massive Majorana fermions
\be
{\cal H}_{QAT} (x) = \sum\limits_{a=1,2}{\cal H}^{(m)}_M [\xi_a (x)]+
8Ka_0\xi_{1R}(x)\xi_{2R}(x)\xi_{1L}(x)\xi_{2L}(x),
\label{Hds}
\ee
which is obviously equivalent to the massive Thirring model for a single
Dirac fermion. Using the standard rules of Abelian bosonization 
which are briefly summarised in Appendix A.1,
one maps the QAT model onto a $\beta^2 = 4\pi$ quantum SG model
with a marginal perturbation \cite{Coleman}:
\be
{\cal H}_{QAT} \Rightarrow {\cal H}_0 [\Phi]
- 8 K a_0 \p_x \phi_R \p_x \phi_L
- (m/a_0) \cos \sqrt{4\pi} \Phi.
\label{QAT-SG+marg}
\ee

Notice that, the bosonization of the QAT model can be also achieved by 
first mapping the model onto a single chain of spinless fermions, 
and then bosonizing the latter. This alternative route is traced 
in Appendix \ref{Lattice-fermions}.

Turning to the DQAT model, we observe that the $h$-term in (\ref{DQAT})
is nonlocal in terms of lattice Majorana operators. However, 
it is known that, for two identical Ising models close to criticality,
products of
two Ising operators belonging to different chains, 
$\s_1 \s_2,~\mu_1 \mu_2,~ \s_1 \mu_2,~ \mu_1 \s_2$,
all with scaling
dimension 1/4, can be expressed {\sl locally} in terms of bosonic
vertex operators with the same scaling dimension (see, for instance, \cite{GNT}
and references therein). 
In Appendix A.2, we rederive this correspondence
starting from the lattice theory of two quantum Ising chains
and paying special attention to the Klein factors:
\bea
\mu^z _{1,n+1/2} \mu^z _{2,n+1/2} &=&  :\cos \sqrt{\pi} \Phi(x):,
\label{mumu1}\\
\s^z _{1n} \s^z _{2n} &=&
- \ri \kappa_1 \kappa_2 :\sin \sqrt{\pi} \Phi (x):,
\label{ss1}\\
\s^z _{1n} \mu^z _{2,n+1/2} &=&
- \ri \kappa_1 :\cos \sqrt{\pi} \Theta(x):,
\label{smu1}\\
\mu^z _{1,n+1/2} \s^z _{2n} &=&
\ri \kappa_2 :\sin \sqrt{\pi} \Theta(x):.
\label{mus1}
\eea
Here $\Theta = - \phi_R + \phi_L$ is a scalar field dual to $\Phi$. 

Thus, Eq. (\ref{ss1}) establishes the continuum bosonized version
of the ``magnetic field'' coupling term in (\ref{DQAT}):
\begin{equation}
h \sum_n \s_{1n}^z  \s_{2n}^z =  \ri \left( \frac{h}{a_0} \right)
\kappa_1 \kappa_2 \int \rd x~ :\sin \sqrt{\pi} \Phi (x) :.
\label{hterm}
\end{equation}
Notice that algebra (\ref{Klein-algebra}) of the Klein factors allows one
to identify them with Pauli matrices,
$$
\kappa_1 = \tau_1, ~~\kappa_2 = \tau_2, ~~ 
\kappa_1 \kappa_2 = \ri \tau_3.
$$
Therefore, in the continuum limit, the DQAT model can be represented in a
diagonal 2$\times$2 matrix form:
\bea
\lim_{a_0 \rightarrow 0} H_{DQAT} &=& \int \rd x~ \hat{\cal H} (x),
\nonumber\\ 
\hat{\cal H} (x) &=&
\left(
\begin{array}{clcr}
{\cal H}_+ (x)&& 0\\
0&& {\cal H}_- (x)
\end{array}
\right) ,
\label{matrixDGS}
\eea
with the two Hamiltonians
\bea
{\cal H}_{\pm} (x) &=& {\cal H}_0[\Phi(x)]
-8 K a_0 \p_x \phi_R(x)\p_x\phi_L(x)\nonumber\\
&-& \left( \frac{m}{a_0} \right)\cos\sqrt{4\pi}\Phi(x)
 \mp \left( \frac{h}{a_0} \right) \sin\sqrt{\pi}\Phi(x)\, ,
\label{QAT-DSG}
\eea
having the structure of the DSG model (\ref{HDSGbis}) 
and differing only in the sign of the coupling constant $h$.
The identification of the parameters is as follows:
$\gamma=8 K a_0$, $g=m/a_0$, and $\lambda=
h/a_0$ (it is understood that when $a_0\to 0$, $K \to \infty$,
$m,h \to 0$ so as to keep the DSG parameters finite).

The  2$\times$2 matrix form of $\hat{\cal H}$ reflects the 
symmetry of the DQAT model 
with respect to the interchange 
of the two constituent quantum chains, (${\cal P}_{12}$),
under which the DSG Hamiltonians ${\cal H}_{\pm}$ transform to each other.
Formally, this symmetry appears as ``unphysical'' because the Hamiltonian 
$\hat{\cal H}$ commutes with $\tau_3$.
This allows one to set
\be
\kappa_1 \kappa_2 = \ri, \label{proj-H+}
\ee
and 
study the Ising transition by
projecting the matrix DQAT model onto the subspace
of a single DSG Hamiltonian ${\cal H}_+$
(the choice $\kappa_1 \kappa_2 = -\ri$, ${\cal H} \rightarrow {\cal H}_-$
would be as good as the above one). However, while 
$\kappa_1\kappa_2$ is a conserved quantity, each $\kappa_i$ is not, 
implying that correlations functions of operators which contain 
the Klein factors may involve transitions between the two different 
sectors $\tau_3=\pm 1$. 

We shall see below that some physical models are automatically mapped onto
a single DSG model with a fixed sign of $h$ (section 5),
while some others reduce to the matrix form
(\ref{matrixDGS}) (section 6). 
In the latter case,
the effective ${\cal P}_{12}$ symmetry corresponds to
a {\sl physical} discrete $Z_2$ symmetry of
the model in question, which is spontaneously broken in the ground state,
a typical example of this kind being a spontaneously dimerized phase. 
In such situations, the Hamiltonians
$H_{\pm}$, when considered separately, describe only
excitations above each of the two degenerate vacua. On the other hand,
the two-fold degeneracy of the ground state necessarily implies
the existence of topological kinks
which interpolate between neighbouring degenerate minima of 
two {\sl different} potentials ${\cal U}_{\pm}$, corresponding to
${\cal H}_{\pm}$. However, 
in the entire parameter space of the model (\ref{matrixDGS}),
including the Ising transition point, the sets of minima of the potentials
${\cal U}_{\pm}$ always have a finite relative shift. This means that
the topological $Z_2$ kinks always remain massive and 
thus appear as irrelevant excitations as long as Ising criticality
is concerned.

Thus, when working with a single Hamiltonian ${\cal H}_+$,
all operators 
representing strongly fluctuating fields at the Ising critical point
are automatically taken into account. All ``off-diagonal'' operators 
proportional
to $\kappa_1$ or $\kappa_2$, such as those given by (\ref{smu1}) and 
(\ref{mus1}),
can be dropped as they represent short-ranged fields
at the critical point.

\section{Deformed quantum Ashkin-Teller 
model in the $(\s,\tau)$ representation
and the Ising transition}
\label{DAT}

The original two-chain representation of the Hamiltonian (\ref{DQAT}) 
is not the most appropriate one to describe the role of the 
coupling term (\ref{hterm}), but some general 
statements can already be drawn at this stage. First of all it is 
clear that, if the chains are originally in the ordered phase, 
the role of the $h$-term would be exhausted by removing the
degeneracy between the four ground states specified by
the signs of the order parameters $\sigma_1 = \langle\sigma^z_{1n}\rangle$ 
and $\sigma_2 = \langle \sigma^z_{2n}\rangle$. The lowest-energy sector,
determined by the condition $h\sigma_1\sigma_2>0$, will stay massive anyway.
(For two ordered Ising copies, the $h$-term gives rise to an effective
longitudinal magnetic field applied to both Ising systems and
thus keeping them massive.)
Therefore, new effects can be only expected if
the two chains are originally disordered.

We can give an heuristic argument in favour of an Ising critical point that 
appears at a finite $h$, if at $h=0$ the chains are disordered.  
For the sake of clarity, let us make a duality transformation and
replace the original model by a pair of two ordered Ising copies well
below $T_c$, coupled by the interaction $h \mu_1 \mu_2$.
Consider a single quantum Ising model (\ref{QI}) at $J \gg \Delta$.
In the leading approximation, its ground state is fully polarised.
The disorder operator $\mu^z _{n+1/2}$ creates a domain wall by flipping
all the spins within the interval $1\leq j \leq n$. 
At short distances, the domain walls
behave like hard-core bosons which, in the dilute limit, are equivalent to 
spinless fermions with the dispersion
$$
\epsilon (k) = 2 J - 2 \Delta \cos k a_0.
$$
The coupling term $h \mu_1 \mu_2$ creates or destroys a pair of domains
walls, one for each chain. Therefore, in the dilute limit, the effective
Hamiltonian for the fermions describing domain walls is of the form:
\be
H = \sum_{k} \sum_{i=1,2}\epsilon (k) c^{\dagger}_{i,k} c_{i,k} 
+ h_{\em eff} 
\sum_k \left( c^{\dagger}_{1,- k} + c_{1,k} \right)
\left(c^{\dagger}_{2, k} + c_{2,- k} \right)
\label{dilute-ham} 
\ee
where $ h_{\em eff} \propto \Delta h/J$.
The combinations of the creation and annihilation operators that
appear in the coupling term reflect the fact that 
$\left[\mu^z _{n+1/2}\right]^2 = 1$.
It is clear that the model (\ref{dilute-ham}) is a theory of four Majorana
fermions, two of which are not affected by the coupling term.
The remaining part of Hamiltonian can be straightforwardly diagonalized
and reduced to two Majorana fields with the masses
$$
m_{\pm} = 2J \pm {\rm const.} \frac{\Delta h}{J}. 
$$
The Ising criticality is reached when one of the masses vanishes, which
occurs when $h \sim J^2 /\Delta$. In the dual representation, the above
estimation of the critical field should be replaced by
$h \sim \Delta^2/J $.
Strictly speaking, the domain walls do not posses a fermionic
statistics but contain a Jordan-Wigner type phase, which
can only be neglected in the extreme dilute limit.

Although the above argument provides a 
qualitative explanation of the transition,
a satisfying description of the critical region can only be reached 
if we are able to properly identify the degrees of freedom 
which get critical and those which do not
(and, of course, take into account the correct stastistics
of the domain walls).    
For this purpose, let us introduce a new Ising variable  
\be 
\tau^z_n=\s^z_{1,n}\s^z_{2,n}.
\label{relation}
\ee
Namely, let us switch from the original two-chain representation, 
with basic spin operators $\s^z _{1n}$ and $\s^z _{2n}$, 
to a new one where the basic variables
are $\tau^z_{n}$ and $\s^z _{n} = \s^z _{1n}$ (due to the ${\cal P}_{12}$
symmetry of the model, the choice $\tau^z_{n}$ and $\s^z _{n} = \s^z _{2n}$
would be equivalent). In the original ($\s_1$-$\s_2$) representation,
the local (i.e. at a fixed lattice site $n$) Hilbert space of the two-chain
model is spanned by the basis vectors  $|\s_1, \s_2 \ra$ which are eigenstates 
of
$\s^z _{1}$ and $\s^z _{2}$:
$$
\s^z _{1,2} |\s_1, \s_2 \ra = \s_{1,2} |\s_1, \s_2 \ra,
~~~\s_{1,2} = \pm 1.
$$
The new local basis $|\s, \tau \ra$ is defined as
$$
\s^z |\s, \tau \ra = \s |\s, \tau \ra, ~~
\tau^z |\s, \tau \ra = \tau |\s, \tau \ra ,
$$
where $\s = \s_1,~\tau =\s_1\s_2$. Comparing matrix elements of the
operators $\s^{\alpha}_{1n}$ and $\s^{\alpha}_{2n}$ in the two bases, 
we find the following correspondence:
\bea
\s^z _{1n} = \s^z _n, &~& \s^z _{2n} = \s^z _n \tau^z _n, \nonumber\\
\s^x _{1n} = \s^x _n \tau^x _n, &~& \s^x _{2n} =  \tau^x _n.
\label{s1s2-st}
\eea  

We also need to define the variables $\mu_n$ and
$\nu_n$ dual to $\s_n$ and $\tau_n$, respectively. The pairs
$(\s_n, \mu_n)$ and $(\tau_n,\nu_n )$ should obey the duality relations
(\ref{dual}) and (\ref{dualinv}). Using these relations together with
(\ref{s1s2-st}), one finds out how the dual spins transform under 
the change of basis:
\bea
\mu^z_{1,n+1/2} =\mu^z _{n+1/2} \nu^z _{n+1/2},
&~& \mu^z_{2,n+1/2} =\nu^z _{n+1/2},\nonumber\\
\mu^x_{1,n+1/2}=\mu^x _{n+1/2}, 
&~& \mu^x_{2,n+1/2}=\mu^x _{n+1/2}\nu^x _{n+1/2}.
\label{mu1mu2-munu}
\eea

In the $\s$-$\tau$ representation the DQAT model (\ref{DQAT}) transforms to
another two-chain model which we call the $\s$-$\tau$ model:
\be
H[\s,\tau] = H_{\s} + H_{\tau} + H_{\s\tau}. \label{H-new-rep}
\ee
Here $H_{\s}$ is a QI Hamiltonian similar to (\ref{QI}) but with different 
parameters:
\be
H_{\s} = \sum_n \left( - J \s^z _n \s^z _{n+1}  + K \s^x _n \right).
\label{H-sigma'}
\ee
$H_{\tau}$ is also of the QI type model but the magnetic field is nonzero
both in the transverse and longitudinal directions:
\be
H_{\tau} = K \sum_n \tau^z _n \tau^z _{n+1}
-  \sum_n \left( h \tau^z _n + \Delta \tau^x _n  \right).
\label{H-tau'}
\ee 
Finally, the coupling term is of the Ashkin-Teller type:
\be
H_{\s\tau} = - \sum_n \left( J \s^z _n \s^z _{n+1} \tau^z _n \tau^z _{n+1}
+ \Delta \s^x _n \tau^x _n  \right).
\label{H-sigma-tau'}
\ee

A mean-field approach to (\ref{H-new-rep}) is outlined
in Appendix \ref{meanfield}, here we shall concentrate
on the large-$h$ limit. 
For large $h$, the $\tau$-degrees of freedom freeze in a 
configuration where $\langle \tau^z_n\rangle \simeq 1$, and 
$\langle \tau^x_n\rangle \simeq \Delta/h$. The 
$\sigma$-degrees of freedom are then described by an effective 
Ising model 
\be
H[\s] \rightarrow - \sum_n \left( 2 J \s^z _n \s^z _{n+1} 
+ \left(\frac{\Delta^2}{h}-K\right) \s^x _n \right),
\label{Hseffect}
\ee
which 
can indeed become critical when $\Delta \sim \sqrt{Jh}$. Although 
this simple picture holds only when $h\gg \Delta \gg J,K$, 
if a universal behaviour is to be expected, then we are lead to conclude 
that, both at strong and weak $h$, the Ising transition essentially 
corresponds the situation when the $\sigma$ degrees of freedom go massless,
while the $\tau$ degrees of freedom remain frozen in a disordered 
configuration with both $\la \tau^z \ra$ and $\la \tau^x \ra$ 
nonzero.
Notice that, with the above strong-coupling description, 
we are unable to determine 
the critical line and even estimate the 
strength of the {\sl irrelevant} operators close to this line.   
This means that, even though the universal properties of the DSG model 
at the Ising transition, including the singular
parts of physical quantities and critical exponents of the correlation
functions, will be captured correctly, the nonuniversal parts, such as 
prefactors and subleading corrections to main asymptotics,
are not to be trusted. 

Let us conclude this section by noticing two 
important facts which will prove useful 
in the next Section devoted to the correlation functions.
\begin{itemize}
\item[{\bf (i)}]
Quantum critical points are associated with gapless phases which are
realized under certain conditions imposed on the parameters of
the model. As soon as any of those parameters
is shifted away from the criticality constraint,
the system becomes off-critical. The corresponding perturbations
to the critical Hamiltonian are therefore
{\sl relevant} operators with conformal dimensions
determined by the universality class of the critical model.
Notice that there is no magnetic field coupled to $\s^z$ in the exact
lattice Hamiltonian (\ref{H-new-rep}). Therefore a departure from
criticality by changing any one of the couplings ($h$, $J$,
$\Delta_J$,
or $K$) will give rise to an Ising mass term.

\item[{\bf (ii)}]
From Eq.(\ref{Hseffect}), if we reasonably take $|K|<2J$, we arrive at 
the conclusion that, for $h>h_c$, the $\sigma$ degrees of freedom 
are ordered, while they are disordered otherwise.
\end{itemize}

\section{Correlation functions}
\label{corf}

In this Section we use the results of the above
quantum-spin-chain type mappings to determine the
physical properties of the DSG model close to
the criticality. 

\subsection{DSG operators in 
the $(\s,\tau)$ representation: UV-IR transmutation}

In preceding sections we have shown that (i) the DQAT
model can be mapped onto the DSG model which is a Gaussian
free theory of the field $\Phi(x)$ in the ultraviolet
(UV) limit, and that (ii) the DQAT can also be 
mapped onto the $(\s,\tau)$ model which, at the Ising 
transition,
essentially reduces to a single critical Ising model
of the order field $\s(x)$ and disorder field
$\mu(x)$
in the infrared (IR) limit. Our aim here is to find out 
how the operators of the DSG model, originally defined in the 
vicinity of the UV fixed point, ``transmute'' when going from the
UV limit to the IR limit.

\subsubsection{Current operators}

We start with holomorphic, or current, operators that are made up of additive
(analytic and anti-analytic) chiral parts. Of physical interest are 
the vector and axial current densities which, in terms of the bosonic field
of the DSG model, are defined as
\bea
J(x) &=& J_R (x) + J_L (x) = \frac{1}{\sqrt{\pi}} \p_x \Phi (x), \label{1cur}\\
J_5 (x) &=& J_R (x) - J_L (x) = - \frac{1}{\sqrt{\pi}} \p_x \Theta (x)
= - \frac{1}{\sqrt{\pi}} \p_t \Phi(x) .
\label{1cur5}
\eea
In physical situations, $J(x)$ determines the smooth part of the charge or
spin density [$J(x) \equiv \rho (x)$], while $J_5 (x)$ describes the 
corresponding charge or spin current [$J_5 (x) \equiv j(x)$].
As follows from (\ref{cur-R}), (\ref{cur-L}), these can be expressed in
terms of the Majorana fields $\xi^1$ and $\xi^2$:
\bea
J(x) &=& \ri \left[ \xi_{1R}(x) \xi_{2R}(x)
 +  \xi_{1L} (x)\xi_{2L}(x) \right], \label{2cur}\\
J_5 (x) &=&  \ri \left[ \xi_{1R}(x) \xi_{2R}(x)
 -  \xi_{1L} (x)\xi_{2L}(x) \right]. \label{2cur5}
\eea
Making the inverse chiral rotation from $(\xi^a _R, \xi^a _L)$ to 
$(\eta^a,\zeta^a)$, $~(a=1,2)$,
we can define a local lattice operator
\be
J_n = \frac{\ri}{2} \left( \eta_{1n} \eta_{2n} + \zeta_{1n} \zeta_{2n}
\right),
\ee
which reproduces (\ref{2cur}) in the continuum limit. 
Using the inverse Jordan-Wigner relations (\ref{Majbis}) and transformations
(\ref{s1s2-st}),(\ref{mu1mu2-munu}), we obtain:
\bea
J_n &=& - \frac{\ri}{2} \kappa_1 \kappa_2 \s^z _{1n}\s^z _{2n}
\left( \mu^z _{1,n+1/2} \mu^z _{2,n+1/2} -
 \mu^z _{1,n-1/2} \mu^z _{2,n-1/2}\right) \nonumber\\
&=& \frac{1}{2}  \tau^z _{n}
\left( \mu^z _{n+1/2} - \mu^z _{n-1/2} \right)
\label{rho_n-via-s-tau}
\eea
(here we have implemented our Klein factor convention (\ref{proj-H+})).
Using the fact that the $\tau$-field is noncritical and has a nonzero
expectation value, we pass to the continuum limit and thus find the expression
for the current density at the infrared fixed point:
\be
J (x) \to C \p_x \mu (x),
\label{cur-IR}
\ee
where $C \sim \la \tau^z\ra$ is a nonuniversal number,
and $\mu (x)$ is the Ising disorder field
at the criticality. Thus, the UV-IR transmutation of the current density
is given by
\be
J(x)=
\left\{ 
\displaystyle
\begin{array}{ll}
\frac{1}{\sqrt{\pi}}\p_x\Phi(x) & {\rm UV},\\
C\p_x\mu(x) & {\rm IR}.
\end{array}
\right.
\label{csum}
\ee 

Turning to the axial current $J_5$, we notice that 
the latter is related to
the vector current $J$ via the continuity equation:
\be
\p_t J(x,t) + \p_x J_5 (x,t) = 0. \label{cont.eq}
\ee
As a result, the IR form of $J_5(x)$ can be immediately recovered. Indeed,
the Ising disorder field $\mu$ is a  scalar field (with zero conformal spin).
This means that holomorphic properties of the vector and axial current operators 
are lost at the IR fixed point. Then the result
(\ref{cur-IR}), together with the 
requirements of Lorentz invariance and continuity equation (\ref{cont.eq}),
leads to the correspondence
\be
J_5 (x)=
\left\{ 
\displaystyle
\begin{array}{ll}
- \frac{1}{\sqrt{\pi}}\p_t\Phi(x) & {\rm UV},\\
- C\p_t\mu(x) & {\rm IR},
\end{array}
\right.
\label{c5sum}
\ee
with the same prefactor $C$ as in Eq.(\ref{csum}), provided that
the velocity is set $v = 1$.

As an important consistency check, we still need to explicitly 
determine the lattice version of the current operator. 
One might na\"ively conclude that, since under the inverse chiral rotation
(\ref{chiral-rot2}) the axial current density transforms to
\be
J_5 (x) = - \ri \left[\zeta_1 (x) \eta_2 (x)  + \eta_1 (x) \zeta_2 (x) 
\right],
\label{J-5-eta-zeta}
\ee
it would be sufficient to replace the r.h.s. of (\ref{J-5-eta-zeta})
by its local (single-site) counterpart which one might identify with
$J^5 _n$. This route is misleading because, in the lattice formulation,
the particle current 
is always defined on a {\sl link} $(n,n+1)$ and, therefore,
should be determined from the equation of motion
\be
\ri \p_t J_n = \left[J_n, H_{DQAT}\right].
\label{contl}
\ee
which, in the continuum limit, is supposed to reduce to Eq.(\ref{cont.eq}).
It can easily be checked that the (density) operator, $J_n$, 
commutes with all the interaction terms in the lattice
DQAT Hamiltonian (\ref{DQAT}), since all of them, including the $h$-term, 
are made up of the density operators. Therefore the above
commutator is only contributed to by the Majorana kinetic energy term in
(\ref{DQAT}) and so is easily computed:
\be
\ri \p_t J_n = Q_{n,n+1} - Q_{n-1,n},
\ee
where
\be
 Q_{n,n+1} = - \frac{J}{4} \left(\eta_{1,n+1} \zeta_{2n} + 
\zeta_{1n} \eta_{2,n+1} \right).
\ee
In terms of the lattice spins this reads
\be
J^5_{n,n+1}=-\frac{J}{2}(\tau^z_n+\tau^z_{n+1})\mu^y_{n+1/2}.
\label{jlattice}
\ee
In the IR continuum limit this becomes $\sim \p_t\mu$
since $\p_t\mu\sim i[\mu^z,H_{QI}]\sim \mu^y$, 
as immediately follows
from the dual version of the QI spin chain Hamiltonian.
Thus we arrive at the IR representation of the axial current given by
(\ref{c5sum}).

\subsubsection{Vertex operators}

The analysis of the vertex operators
\[
V_{\beta}[\Phi]=e^{i\beta\Phi}
\]
is in fact simpler than that of the holomorphic
operators. There are two cases of physical interest:
$\beta=\pm\sqrt{\pi}$
and $\beta=\pm \sqrt{4\pi}$.

From (\ref{ss1}), (\ref{proj-H+}) and (\ref{relation}) it follows
that
\be
\sin \sqrt{\pi} \Phi \sim  \s_1 \s_2 \sim  \tau \sim I,
\label{sin}
\ee
($I$ being the identity operator), as the $\tau$ model
is always off-critical.
 This is a reasonable result 
since the above operator is directly present in the
DSG Hamiltonian (\ref{HDSGbis}).
However, as we mentioned in the previous Section,
the operator $\sin\sqrt{\pi}\Phi$ corresponds to
the departure from the criticality of the
$\sigma$-model.
So, this operator, though having a finite average
value, should also possess an extra term 
which, at the critical point, represents a strongly
fluctuating field (with a power-law decaying correlation function). 
A more correct version of the formula (\ref{sin})
is therefore as follows:
\be 
\sin\sqrt{\pi}\Phi \sim I+ \varepsilon,
\label{sincor}
\ee
where $\varepsilon$ is the energy density 
(or a Majorana mass bilinear) operator.
This follows from the first observation {\bf (i)}
at the end of Section \ref{DAT}, which states that moving $h$ 
from its critical value 
results in the stress-energy
tensor renormalisation and, more 
importantly, in the appearance of the
Majorana mass, i.e. the Ising energy-density operator.

Furthermore, (\ref{mumu1}) and the 
lattice fusion rule (\ref{mu1mu2-munu}) give 
\be
\cos\sqrt{\pi}\Phi\sim\mu_1\mu_2\sim\mu.
\label{cos}
\ee
Thus, the operator $\cos\sqrt{\pi}\Phi$ is the most divergent operator
of the DSG model, with a nonzero expectation value at $h < h_c$ and
vanishing upon approaching the Ising critical point as
\be
\la \mu \ra \sim \left( h_c - h  \right)^{1/8}.
\label{av-mu-crit}
\ee

Finally, the behaviour of $\beta=\pm \sqrt{4\pi}$
operators
is determined in full analogy to the above discussion:
\be
V_{\pm\sqrt{4\pi}}[\Phi]\sim I+\epsilon.
\label{4pi}
\ee

Our results on UV-IR DSG operators transmutations are
summarised in table I. 

\begin{table}
\caption{UV-IR transmutation of DSG-operators}
\begin{center}
\begin{tabular}{|c|c|c|c|}
\hline UV limit & $\p_x\Phi$ &
$V_{\sqrt{\pi}}+V_{-\sqrt{\pi}}$
& $V_{\sqrt{\pi}}-V_{-\sqrt{\pi}}$,
$V_{\pm\sqrt{4\pi}}$ \\ 
\hline IR limit & $\p_x\mu$ & $\mu$ & $I+\epsilon$\\ 
\hline
\end{tabular}
\end{center}
\end{table}

\subsection{Correlation functions}

Having identified the operators at the IR fixed point,
we can now determine leading asymptotics of the correlation functions.
Most of the operators of physical interest
can be expressed in terms of the Ising disorder
operator $\mu$ at the Ising transition point 
and in its close vicinity. 

For physical applications of the DSG model, it is
important to investigate the dynamical susceptibility 
defined as the frequency-momentum Fourier transform
of the retarded auto-correlation function of the Ising
disorder parameter:
function
\be
D^{(R)}(\omega,p)= -i\int_{-\infty}^\infty dx
\int_0^\infty dt e^{-ipx+i\omega t}
\la\left[\mu(x,t),\mu(0,0)\right]\ra.
\label{ret}
\ee
It is known that,
at criticality, $D(r)= 1/r^{1/4}$ where 
${\bf r}=(\tau,x)$ ($\tau =\ri t$, 
and $v$ is set to $1$). Furthermore,
away from criticality, in the
{\it ordered} phase, $D(r)=(A_1/\pi)K_0(mr)$
where $A_1$ is related to the Glaisher
constant. The crossover between these two regimes,
$r\ll \xi=1/m$ and $r> \xi$, respectively, 
is complicated and is described in terms
of the Painlev\'{e} theory \cite{MC}. 
If, following \cite{MC}, we introduce,
\[
\zeta=r \frac{d\ln D}{dr},
\]
then the function $\zeta$ can be shown to satisfy
\[
(r\zeta'')^2=
4(r\zeta'-\zeta^2)(r\zeta'-\zeta)+(\zeta')^2,
\]
which is related to Painlev\'{e} V equation and can,
in turn,
be shown to produce the correct conformal and massive
limits.
As the exact expressions are uncomfortable 
for calculations, we shall estimate the
Fourier transforms from the asymptotes of the
correlation function $D(r)$.
 Therefore, first we summarise known
results on the limiting behaviour of this function.

At criticality
\be
D_0(r)=\frac{1}{r^{1/4}}.
\label{Dnot}
\ee
It is understood that this asymptotics is still valid 
away from criticality in the region $r\ll\xi$.

In the ordered phase 
\be
D_>(r)=\frac{A_1}{\pi}K_0(mr),
\label{Dord}
\ee
the large-$r$ limit of which is
\be
D_>(r)=\frac{A_1\sqrt{\pi}}{\sqrt{2m}}\frac{1}{r^{1/2}}
e^{-mr}.
\label{Dordas}
\ee

In the disordered phase, thanks to
the work by McCoy, Wu, Tracy, and collaborators \cite{MC},
we know that
\bea
D_<(r)&=&\frac{A_1}{\pi^2}\left\{
m^2r^2\left[K_1^2(mr)-K_0^2(mr)\right]-mrK_0(mr)K_1(mr)
\nonumber\right.\\&+&\left.
\frac{1}{2}K_0^2(mr)\right\},
\label{Ddis}
\eea
which is understood to be the connected part of the
correlator
(i.e. without the constant piece). The large-$r$
asymptotics
of this function can be obtained from the relevant expansions
of the modified Bessel functions quoted e.g. in \cite{GNT}:
\be
D_<(r)=\frac{A_1}{8\pi m^2}
\frac{1}{r^2}e^{-2mr}.
\label{Ddisas}
\ee

The retarded function (\ref{ret}) can be
found by calculating the Fourier transforms of
the above asymptotic forms with a subsequent
analytic continuation to real frequencies.
This procedure is outlined in Appendix \ref{corrf}.
The results are as follows.

\begin{itemize}
\item{\bf Criticality.} Using (\ref{fan})
and (\ref{Dnotq}), one obtains
\bea
D_0^{(R)}(\omega,p)&=& 2^{1/4}\pi
\frac{\Gamma(7/8)}{\Gamma(1/8)}
\left\{
\frac{\theta(p^2-\omega^2)+\cos(7\pi/8)
\theta(\omega^2-p^2)}{|\omega^2-p^2|^{7/8}}
\right. \nonumber \\   &-&
\left.\frac{2i\sin(7\pi/8)\theta(\omega^2-
p^2)}{|\omega^2-p^2|^{7/8}} \right\}.
\label{Dnotfin}
\eea
\item{\bf Ordered Phase.} Here 
we have a particle pole (\ref{Dorq}):
\be
D_>^{(R)}(\omega,p)=
\frac{\pi A_1}{-\omega^2+\epsilon_p^2}-
\frac{i\pi^2 A_1}{2\epsilon_p}
\left[\delta(\omega+\epsilon_p)+\delta(\omega-\epsilon_p)
\right].
\label{Dorfin}
\ee
\item{\bf Disordered phase.} Specialising to $\alpha=-1/2$ 
in (\ref{fan}) we have:
\be
f(\omega)=
\sqrt{|\omega^2-\epsilon_p^2|}\theta(\epsilon_p^2-\omega^2)
+2i
\sqrt{|\omega^2-\epsilon_p^2|}
\theta(\omega^2-\epsilon_p^2).
\label{fanbis}
\ee
Therefore, as long as $\omega^2<p^2+4m^2$,
the argument of the log-function in (\ref{Ddisq})
is real and positive, so the correlation function
remains real without dissipation.
Above the two-particle threshold
($\omega^2>p^2+4m^2$),
the correlation function has a branch cut and the
dissipative part will appear as follows:
\bea
&~&D_<^{(R)}(\omega,p)=\nonumber\\
&~&\frac{A_1}{8m^2}\left\{
\ln\left(\frac{4m}{
\sqrt{|-\omega^2+p^2+4m^2|}+2m}\right)
\theta(p^2+4m^2-\omega^2)\right.\nonumber\\
&~&-i\left.{\rm
Arctg}\frac{\sqrt{\omega^2-p^2-4m^2}}{m}
\theta(\omega^2-p^2-4m^2)\right\}.
\label{Ddisfin}
\eea
\end{itemize}
The excitations are therefore incoherent in this regime.

\section{Dimerized Heisenberg chain in 
a staggered magnetic field}

The DSG model exhibiting a nontrivial flow towards Ising criticality
can be realized as an effective continuum theory for 
a number of quantum 1D models of strongly correlated
electrons, in particular quantum spin chains and ladders.
In the context of spin systems, an effective DSG model
can emerge within the Abelian bosonization scheme
when staggered fields
breaking translational invariance, such as an explicit dimerization
(bond alternation)\cite{aff-conf} or a staggered magnetic field,
are added to an originally translationally invariant model with a gaped
ground state\footnote{While alternation of the nearest-neighbour
exchange constants can originate from the
spin-phonon coupling, 
the case of a nonuniform magnetic field with 
a period $2a_0$ ($a_0$ being the lattice constant)
used to be regarded as unrealistic, not achievable in experimental conditions.
Fortunately, the status of the staggered magnetic field has recently changed
from exotic to legitimate.
It has been shown that, in some quasi-1D antiferromagnetic compounds,
such field can be realized as an intrinsic one. An effective 
staggered field can
originate from the Neel ordering of one magnetic sublattice and being
experienced by loosely connected magnetic chains which form another sublattice
and remain disordered down to very low temperatures \cite{MZ}.
A sign-alternating component of the magnetic field can be also 
effectively generated
due to a staggered anisotropy of the gyromagnetic tensor,
as it is the case for the Copper-Benzoate organic molecule \cite{benzoate}.}.

Perhaps the simplest example of this kind is given by
the spin-1/2 Heisenberg chain with nearest-neighbour ($J_1$) and
next-nearest-neighbour ($J_2$) antiferromagnetic exchange interactions
\be
H_{J_1-J_2} = \sum_n \left(J_1 {\bf S}_n \cdot {\bf S}_{n+1}
+ J_2 {\bf S}_n \cdot {\bf S}_{n+2}  \right)
\label{J1J2lattice}
\ee

This model
has been extensively studied during past years. If frustrating interaction
$J_2$ is small enough, the model maintains the critical properties of the 
unfrustrated Heisenberg chain ($J_2 = 0$). 
At $J_2 \geq J_{2c}\simeq 0.24 J_1$, frustration
gets relevant and drives the model to a massive phase characterised by
spontaneously broken parity\cite{j1j2,WA}. The ground state is dimerized and 
doubly
degenerate, and there exist massive elementary excitations - topological
$Z_2$ kinks carrying the spin 1/2. At a special (Majumdar--Ghosh)
point, $J_2 = 0.5 J_1$, the picture is particularly simple because 
the two $Z_2$-degenerate ground states are given by
matrix products of singlet dimers formed either on the lattice
links $<2n,2n+1>$ or $<2n-1,2n>$.

Under assumption that $J_1 \gg J_2$, the continuum limit of the
$J_1 - J_2$ model can be considered, and the resulting quantum field theory
is that of a critical $SU(2)_1$ Wess--Zumino--Novikov--Witten
(WZNW) model perturbed a marginal
current-current interaction\cite{WA}:
\be
H_{J_1 - J_2} = \frac{2\pi}{3} \left( 
:{\bf J}_R \cdot {\bf J}_R: + :{\bf J}_L \cdot {\bf J}_L: \right)
+ \gamma {\bf J}_R \cdot {\bf J}_L,
\label{J1-J2-WZW}
\ee
where $\gamma \sim J_{2c} - J_2 > 0$. At $\gamma < 0$, the perturbation is
marginally irrelevant. However, at $\gamma > 0$ the effective interaction flows
to strong coupling, and the system ends up in a spontaneously dimerized phase
with a dynamically generated spectral gap\cite{WA}:
$m_{\em dim} \sim \Lambda \exp (- 2\pi/\gamma)$, 
where $\Lambda \sim J_1$ is the UV cutoff.

Using Abelian bosonization, we can rewrite (\ref{J1-J2-WZW})
as a sine-Gordon (SG) model:
\be
H_{J_1 - J_2} = \frac{1}{2} \left[ \left( \p_x \Phi \right)^2 + 
\left( \p_x \Theta \right)^2 \right]
+ \frac{\gamma}{2\pi} \p_x \Phi_R \p_x \Phi_L
- \frac{\gamma}{(2\pi \alpha)^2} \cos \sqrt{8\pi} \Phi.
\label{J1-J2-bos}
\ee
The ``hidden'' SU(2) symmetry of this bosonic Hamiltonian
is encoded in the robust structure of the last two terms in 
Eq.(\ref{J1-J2-bos}),
parametrised by a single coupling constant $\gamma$. This fact enforces
the SG model (\ref{J1-J2-bos}) to occur either on the weak-coupling SU(2)
separatrix of the Kosterliz-Thouless phase diagram ($\gamma < 0$), or
on the strong-coupling SU(2) separatrix ($\gamma > 0$). 
In the latter case, quantum solitons with the mass $m_{\em dim}$ and
topological charge $Q = 1$ are identified with the $Z_2$ dimerization
kinks carrying the spin $S = Q/2 =1/2$.
In the remainder of this 
section we will be dealing with the massive, spontaneously dimerized phase.

Consider the following deformation of the model:
$
H = H_{J_1 - J_2} + H',
$
where
\be
H' = \sum_{a=0}^3 \lambda_a Tr \left(\tau_a \hat{g} \right).\label{deform}
\ee
Here $\hat{g}$ is the $2\times 2$ WZNW matrix field with conformal
dimensions (1/4, 1/4), and $\tau_a$ are the Pauli matrices including the
unit matrix $\tau_0 = I$. 
The scalar and vector parts of $\hat{g}$,
\bea
{\bf n}_s &\sim& Tr \left(\vec{\tau}\hat{g}   \right)
\sim \left(\cos\sqrt{2\pi} \Theta, \sin\sqrt{2\pi} \Theta,
-  \sin\sqrt{2\pi} \Phi \right), \label{n-bos}\\
\epsilon_s &\sim& Tr \left(\hat{g}   \right) \sim \cos\sqrt{2\pi} \Phi,
\label{eps-bos}
\eea
constitute the staggered magnetisation and dimerization field of the S=1/2
Heisenberg chain.
Representing
\be
H' = \lambda \epsilon_s + {\bf h}_s \cdot {\bf n}_s, \label{deform1}
\ee
let us consider the two terms in (\ref{deform1}) separately.

  The role of weak explicit (spin-Peierls) dimerization in the spontaneously
dimerized $J_1 - J_2$ spin-1/2 chain has already been addressed by 
Affleck \cite{aff-conf}.
The effective double-frequency-sine-Gordon potential appearing in this
case is different from the one studied in this paper (c.f. Eq.(\ref{HDSG})):
\be
{\cal U}_{\em dimer} =
- \frac{\gamma}{(2\pi \alpha)^2} \cos \sqrt{8\pi} \Phi
+ \lambda \cos \sqrt{2\pi} \Phi. \label{pot-dimer}
\ee
The $\lambda$-term in (\ref{pot-dimer}) removes the degeneracy between 
the neighbouring minima
of the unperturbed potential $- \cos \sqrt{8\pi} \Phi$ (i.e. between the two
degenerate dimerized ground states) and thus leads to confinement of the
solitons. The main physical effect is the spinon-magnon transmutation: 
deconfined
spinons of the frustrated Heisenberg chain, carrying the spin S = 1/2,
form bound states with S = 0 and S = 1, the latter representing coherent
triplet magnon excitations. 

  Let us concentrate on the case of the staggered magnetic field, ${\bf h}_s$:
\be
H' = {\bf h}_s \cdot {\bf n}_s . \label{stag-deform}
\ee
Choosing ${\bf h}_s = h_s \hat{z}$, we arrive at a bosonic model
\bea
H &=& H_0 [\Phi]
+ \frac{\gamma}{2\pi} \p_x \Phi_R \p_x \Phi_L
- \frac{\gamma}{(2\pi \alpha)^2} \cos \sqrt{8\pi} \Phi \nonumber\\
&-& h_s  \sin \sqrt{2\pi} \Phi,
\label{j1j2-DSG}
\eea
in which we recognise the DSG model with the structure (\ref{HDSG}).
From the analysis of the preceding section we conclude that the
spontaneously dimerized chain in a staggered magnetic field has two
phases separated by a quantum critical point at $h_s = h^* _s$.
At $h_s < h^* _s$ a ``mixed'' phase is realized, with coexisting
dimerization $\la \epsilon_s \ra \neq 0$ and staggered magnetization
$\la {\bf n}_s \ra \neq 0$. Notice that, as opposed to the case of uniform
magnetic field that couples to the (conserved) total magnetisation, 
the dependence ${\bf n}_s = {\bf n}_s (h_s)$ shows
no threshold in $h_s$. Dimerization vanishes at the Ising critical 
point $h_s = h^* _s$ and remains zero in the ``pure Neel'' phase,
$h_s > h^* _s$. The critical field $ h^* _s$ can be estimated 
by comparing the dimerization gap 
$m_{\em dim}$
with the gap that would open up at $\gamma \leq 0$:
$
m_h \sim h^{2/3} _s.
$
So the critical staggered field is exponentially small:
$
 h^* _s \sim \left(m_{\em dim}\right)^{3/2}.
$

Since the spin SU(2) symmetry is broken by the (staggered) magnetic field,
the total spin is not conserved, but the spin projection
$S^z$ is. The latter circumstance allows one to identify the
spin $S^z$ of elementary excitations as the topological quantum number 
of the kinks interpolating between the nearest degenerate minima
of the potential ${\cal U}(\Phi)$ in (\ref{j1j2-DSG}).
According to the structure of ${\cal U}(\Phi)$, there will be ``short''
and ``long'' kinks, carrying the spin
$
S^z _{\pm} = \frac{1}{2} \mp \delta,
$
where $\delta = \delta (h_s)$ smoothly increases from 
$\delta  = 0$ at $h_s = 0$ to $\delta = 1/2$ at $h_s \geq h^* _s$.
Therefore, in the mixed phase,
the original massive S = 1/2 spinon splits into two
topological excitations carrying
fractional spins $S^z _{\pm}$. These spins become $S^z = 1$ and
$S^z = 0$ at the Ising transition, and it is just the singlet kink which
loses its topological charge and becomes massless at $h_s = h^* _s$.
The existence of the fractional-spin excitations in the mixed phase
$( h_s < h^* _s)$
is nothing but the spin version of the charge fractionization of
topological excitations found
earlier in 1D commensurate Peierls insulators with broken charge conjugation
symmetry (e.g. cis-polyacetylene)\cite{braz}, and also in a recent study of
a 1D Mott insulator with alternating single-site energy \cite{FGN}.

To estimate the behaviour of physical quantities at the transition, let
us consider an anisotropic ($\gamma_{\parallel}, \gamma_{\perp}$) 
version of the model in which
\bea
H &=& H_0 [\Phi]
+ \frac{\gamma_{\parallel}}{2\pi} \p_x \Phi_R \p_x \Phi_L
- \frac{\gamma_{\perp}}{(2\pi \alpha)^2} \cos \sqrt{8\pi} \Phi \nonumber\\
&-& h_s  \sin \sqrt{2\pi} \Phi,
\label{j1j2-DSG-anis}
\eea
The $\gamma_{\parallel}$-term in (\ref{j1j2-DSG-anis}) can be eliminated
by an appropriate rescaling of the field, $\Phi \rightarrow \sqrt{K} \Phi$:
\be
H \rightarrow H_0 [\Phi] 
- \frac{\gamma_{\perp}}{(2\pi \alpha)^2} \cos \sqrt{8\pi K} \Phi
+ h_s \sin \sqrt{2\pi K} \Phi,
\label{anis-DSG}
\ee
As already mentioned, universality arguments lead to the conclusion
that the anisotropic model (\ref{anis-DSG}) also
incorporates the Ising criticality at some value of $h_s$.
Choosing $K (\gamma_{\parallel}) = 1/2$, we reduce the perturbation to
the form (\ref{HDSGbis})
\be
H' = - \frac{\gamma_{\perp}}{(2\pi \alpha)^2} \cos \sqrt{4\pi}\Phi
+   h_s \sin \sqrt{\pi} \Phi,
\label{desired-DSG}
\ee
discussed in detail in previous sections. Using the ($\s$-$\tau$) 
representation,
we derive the relations in Table \ref{TABB} 
describing the UV-IR transmutation of the physical fields.

\begin{table}
\label{TABB}
\caption{UV-IR transmutation of operators}
\begin{center}
\begin{tabular}{|c|c|c|}
\hline 
field & UV & IR \\
\hline
uniform spin density &  $J^z \sim \p_x \Phi$  & $J^z \sim \p_x \mu$ \\
\hline
uniform spin current & ${\cal J} \sim \p_t \Phi$ &
${\cal J} \sim \p_t \mu$ \\
\hline 
dimerization & $\epsilon_s \sim \cos \sqrt{2\pi} \Phi$ &
$\epsilon_s \sim \mu$ \\
\hline
staggered magnetisation & $n^z \sim \sin \sqrt{2\pi} \Phi$ &
 $n^z \sim I + \varepsilon$\\
\hline
\end{tabular}
\end{center}
\end{table}

We have the following correspondence:
\bea
h_s < h^* _s: && {\em disordered~phase~}: \la \mu \ra \neq 0; \nonumber\\
h_s > h^* _s: && {\em ordered~phase~}: \la \mu \ra = 0. \nonumber
\eea
We see that dimerization is finite at $h_s < h^* _s$ and vanishes as
$$
\la \epsilon_s \ra \sim (h^* _s - h_s)^{1/8},
$$
on approaching the critical point. The staggered magnetisation, on the other 
hand,
is always finite in both phases. Its behaviour at the transition is determined
by the subleading correction to the identity operator (see 
Table \ref{TABB}):
\bea
\la \la n^z \ra \ra &\equiv&  \la n^z \ra_{h_s} - \la n^z \ra_{h^* _s}
\nonumber\\ 
&\sim& \left( h_s - h^* _s  \right) \ln \frac{h^* _s}{|h_s - h^* _s|}.
\eea
The logarithmic divergence of the staggered magnetic susceptibility at the 
transition
is similar to that of the specific heat of the Ising model:
\be
\chi_{stag} \sim \ln \frac{h^* _s}{|h_s - h^* _s|}.
\ee

Next we consider the dynamical magnetic susceptibility.
In analogy to (\ref{ret}), it is defined by
\be
\chi(\omega,p)= -i\int_{-\infty}^\infty dx
\int_0^\infty dt e^{-ipx+i\omega t}
\la\left[S^z(x,t),S^z(0,0)\right]\ra
\label{retm}
\ee

Using the continuum limit decomposition of 
the spin operators and the above glossary,
we conclude that while the uniform magnetic susceptibility
is readily given by 
\be
\chi(\omega, q\sim 0)\sim q^2 D^{(R)}(\omega, q)
\label{chiun}
\ee
(the function $D^{(R)}$ has been extensively
discussed in Section (\ref{corf})),
the staggered susceptibility is in turn related to the
correlation function of the Ising energy-density operator.
The latter object is a Majorana bilinear
\[
\varepsilon(x)\sim \xi_R(x)\xi_L(x).
\]
Therefore, calculating the staggered magnetic susceptibility
reduces to a simple task of computing the polarisation
loop diagram for free, massive Majorana fermions. 
The result is 
\be
\chi(\omega, q=\pi)\sim
\ln\left(1-\frac{\omega^2}{4m^2}
\right)-2\sqrt{\frac{4m^2-\omega^2}{\omega^2}}
{\rm arctg}\left(
\frac{\omega^2}{4m^2-\omega^2}\right),
\label{chiin}
\ee
inside the gap ($|\omega|<2m$) and
\be
\chi(\omega,\pi)\sim
\ln\left(\frac{\omega^2}{4m^2}-1
\right)
-i{\rm sign}\omega\left\{\pi+
2\sqrt{\frac{\omega^2-4m^2}{\omega^2}}
{\rm arctg}\left(
\frac{\omega^2}{\omega^2-4m^2}\right)\right\},
\label{chiout}
\ee
for $|\omega|>2m$.

We notice that for $h>h_s$ it obviously is the
staggered field operator ($\beta^2=2\pi$) which alone
dictates the physics of our effective DSG 
model (not only is it the most relevant
operator, but it also has a large amplitude).
It is therefore instructive to compare
the results for the dynamical magnetic
susceptibility of the DSG model we
have found via Ising-type mappings with
those for the SG model with a $\beta^2=2\pi$
operator only, which were obtained in the
paper \cite{ETD} by means of the
form-factor technique. 
As for $h>h_s$ we enter the ordered phase ($\langle
\mu\rangle$=0), the uniform
susceptibility (\ref{chiun},\ref{Dorfin})
is of a coherent nature, very much
in agreement with \cite{ETD} (`magnon'
contribution). The staggered susceptibility
(\ref{chiin},\ref{chiout}), on the other
hand, is incoherent in excellent agreement
with the kink-antikink continuum observed
in \cite{ETD} (the breather contribution
is missed in our approach; it might be
recovered as a bound state of Majorana
fermions, but we shall not push our analysis
beyond this point).
No analogous comparison can be made for $h<h_s$
(no form-factor calculations are, to our
knowledge, currently available for the 
$\beta^2=8\pi$ SG model, and, even if they were,
a comparison would have been of dubious validity, 
as the addition of the $\beta^2=2\pi$-operator
qualitatively changes the spectrum from the
start).

\section{Ising transition in the 1D Hubbard model with
alternating chemical potential}

The DSG model finds a number of interesting applications in the theory of
1D strongly correlated electron systems. 
In this section we shall consider a particular example of this kind 
which has been recently discussed in Ref.\cite{FGN} -- a 1D repulsive Hubbard 
model
at 1/2-filling with a sign-alternating single-site energy (i.e. staggered 
chemical 
potential). The Hamiltonian of this model reads:
\be
H = - t \sum_{i,\s} \left( c^{\dagger}_{i\s} c_{i+1,\s} + h.c. \right)
+ U \sum_i n_{i\up} n_{i\down}
+ \Delta \sum_{i,\s} (-1)^i n_{i\s},
\label{ham}
\ee
where $c_{i\s}$ is the annihilation operator of an electron with the spin 
projection
$\s$, residing at the lattice site $i$,
and $n_{i\s} = c^{\dagger}_{i\s} c_{i\s}$.
The model (\ref{ham}) was originally proposed in the context of quasi-1D organic 
materials \cite{nagaosa} and is also believed to be prototypical
for ferroelectric perovskites \cite{egami}.

In spite of its apparent simplicity, the model (\ref{ham}) reveals nontrivial
physics. At $U = 0$ it describes a band insulator (BI) with a spectral gap for
all excitations. At $\Delta \ll t$,
the low-energy spectrum of the BI is that of free massive 
Dirac fermions. 
When the Hubbard interaction is switched on,
the finite fermionic mass $\Delta$ makes the theory
free of infrared divergences, so that the BI phase remains stable
for small enough $U$.
On the other hand, at $\Delta = 0$, the Hamiltonian (\ref{ham})
coincides with the standard (translationally invariant) Hubbard model which
is exactly solvable \cite{LW} and is well known to
describe a Mott insulator (MI) at any positive value of $U$,
if the electron concentration $n = (1/N)\sum_{i,\s} n_{i\s} = 1$
(the case of a 1/2-filled energy band). 
The MI state has a finite mass gap $m_c$ in the charge
sector induced by commensurability of the electron density with
the underlying lattice. At energies well below $m_c$, 
local charge fluctuations are suppressed, and the low-energy 
spin dynamics of the model
coincides with that of the spin-1/2 Heisenberg antiferromagnetic chain,
the latter possessing a gapless spectrum.
At a finite $U$, the charge-gaped MI phase is stable against site alternation,
provided that $\Delta$ is small enough \cite{nagaosa}. 

Thus, the issue of interest is the nature of the crossover between the BI and MI
regimes which is expected to occur in the strong-coupling region where the
single-particle mass gap $\Delta$ becomes comparable with the MI charge gap 
$m_c$.
Starting out from the MI phase and decreasing $U$ at a fixed $\Delta$, 
one has to identify the mechanism
for the mass generation in the spin sector. On the other hand, 
it is clear that the charge degrees of freedom should also be involved
in the BI-MI crossover. Indeed,
dividing the lattice into two sublattices, A and B, with the single-site
electron energies $\Delta$ and $-\Delta$, respectively, and
considering
electronic states of a diatomic (AB) unit cell in
the limit $t \ll U, \Delta$, one finds a region $U \sim 2\Delta$ where
two charge 
configurations, $A^1 B^1$ and $A^0 B^2$, become almost degenerate. This is the
so-called mixed-valence regime where those excitations responsible for
the charge redistribution among the two unit-cell configurations become soft.
This means that, apart from the spin transition, a charge transition associated
with vanishing of the charge gap at some value of $U$ is also expected
to occur.

In Ref.\cite{FGN} we have shown that the MI-to-BI crossover, taking place
on decreasing $U$ at a fixed $\Delta$, is realized as a sequence of two
continuous transitions: a Berezinskii-Kosterlitz-Thouless (BKT) transition
at $U = U_{c2}$ where a spin gap is dynamically generated, and an Ising
critical point at $U = U_{c1} < U_{c2}$ where the charge gap vanishes. 
Assuming that $U, \Delta \ll t$,
below we
shall consider the effective low-energy field theory for the lattice model
(\ref{ham}). We shall then briefly comment on the spin transition and
mostly concentrate
on the Ising transition in the charge sector of the model which will be
described in terms of a DSG model. 

The standard bosonization procedure (see e.g. \cite{GNT}) allows one to 
represent
the Hamiltonian density as
$$
{\cal H}_{\em eff} = {\cal H}_c +  {\cal H}_s +  {\cal H}_{cs}.
$$
Here the spin sector is described by the $SU(2)_1$ WZNW model with a
marginally irrelevant current-current perturbation originating from
the electron backscattering processes ($g \sim U a_0 > 0$):
\be
{\cal H}_s = \frac{2 \pi v_s}{3} \left( :{\bf J}_R \cdot{\bf J}_R :
+ :{\bf J}_L \cdot{\bf J}_L : \right)
- 2 g {\bf J}_R \cdot{\bf J}_L \label{spin}
\ee
where ${\bf J}_{R,L}$ are chiral components of the vector spin current 
satisfying the
$SU_1(2)$ Kac-Moody algebra. This Hamiltonian accounts for the 
universal properties of the spin-1/2 antiferromagnetic Heisenberg 
chain in the scaling limit \cite{affleck}. 
The charge degrees of freedom are represented
by a sine-Gordon model for a scalar field $\Phi_c$:
\be
{\cal H}_c = \frac{v_c}{2} \left[ \Pi^2 _c + \left(\p_x \Phi_c \right)^2  
\right]
- \frac{m_0}{\pi \alpha} \cos \sqrt{8\pi K_c} \Phi_c
\label{charge}
\ee
where $m_0 \sim g$. 
The cosine perturbation is caused by 
the electron Umklapp processes (see e.g.\cite{GNT}). For a wide class
of Hamiltonians with finite-range interactions (of which the Hubbard model
is a member) the parameter $K_c  < 1$. This means that the model (\ref{charge})
is in a strong-coupling regime, and the dynamically generated mass
determines the gap $m_c$ in the charge sector.
The charge and spin sectors are coupled by the $\Delta$-term:
\be
{\cal H}_{cs} = - \frac{2\Delta}{\pi \alpha} \epsilon_s
\sin \sqrt{2\pi K_c} \Phi_c 
\label{delta-term}
\ee
where $\epsilon_s$ is the spin dimerization field of the S=1/2 Heisenberg
chain, defined in (\ref{eps-bos}).

Let us assume that we are in the MI regime with a gaped charge sector.
Since the charge field $\Phi_c$ is locked in one of degenerate minima
of the periodic potential in (\ref{charge}),
$$
\left( \Phi_c \right)_m = \sqrt{\frac{\pi}{2K_c}} m, ~~~m \in Z_{\infty}
$$
it follows immediately that, in the MI phase,
the operator $\sin \sqrt{2\pi K_c} \Phi_c$ 
appearing in (\ref{delta-term})
does not represent a strongly
fluctuating field; it is rather short-ranged at distances
$\xi_c \sim v_c / m_c$, and this explains the stability of the MI phase against
a small $\Delta$-perturbation.
Assuming that the charge gap is nonzero, we integrate out the massive
charge degrees of freedom to obtain the effective action in the spin sector.
According to the unbroken SU(2) symmetry of the model, 
the effective Hamiltonian in the spin sector
retains
its form (\ref{spin}), but the current-current coupling constant undergoes
an additive renormalization. In the second order in $\Delta$ we obtain:
$$
g \rightarrow g_{\em eff} = g - C \left(\Delta/ m_c  \right)^2 v_c
$$
where $C \sim 1$ is a nonuniversal numerical constant. The spin sector now
resembles the $J_1 - J_2$ frustrated spin-1/2 chain (see section 5)
with an effecive next-nearest-neighbour interaction $J_2$ generated by
the staggered chemical potential $\Delta$.
As long as $g_{\em eff} > 0$, the spin excitation spectrum
 remains gapless. However, decreasing
$U$ at a fixed $\Delta$ (or increasing $\Delta$ at a fixed $U$) eventually
reverts the above inequality to $g_{\em eff} < 0$. In this case the 
current-current
perturbation becomes marginally relevant, 
and a continuous (BKT) transition takes place to a spontaneously
dimerized insulating (SDI) phase with broken (site) parity and 
a finite mass gap in the spin sector. 

Thus, the condition $g_{\em eff} = 0$ determines the spin transition point 
$U_{c2}$.
Using the exact result for the charge gap in the small-$U$ Hubbard 
model\cite{LW},
$$
m_c \sim \sqrt{Ut} e^{- 2\pi t/U},
$$
we find that
\be
U_{c2} = \frac{2\pi t}{\ln \left(t/U \right)}
\left[1 + O \left(\frac{\ln\ln (t/\Delta)}{\ln (t/\Delta)} \right) \right]
\label{U-c2}
\ee

The dimerization order parameter that becomes nonzero at $U < U_{c2}$
is defined as
\be
D = \sum_{i,\s} (-1)^i \left( c^{\dagger}_{i\s} c_{i+1,\s} + h.c. \right)
\label{dimer-op}
\ee
and in the continuum limit its density is given by
\be
D (x) \sim \cos \sqrt{2\pi K_c} \Phi_c (x) Tr \hat{g}_s (x)
\label{dimer-op-den}
\ee

It is instructive to compare the parity properties of the SDI and BI phases.
For models defined on a 1D lattice, there are two parity transformations --
the site parity ($P_S$) and link parity ($P_L$) \cite{EA}. The difference
between $P_S$ and $P_L$ survives the continuum limit and shows up in
two inequivalent parity transformations that keep the massless Dirac equation
$$
\left(\p_t + \s_3 \p_x  \right) \psi (x) = 0, ~~~ \psi
= \left(
\begin{array}{clcr}
R\\
L
\end{array}
\right)
$$
invariant. These are
\bea
P_S: && \psi (x) \rightarrow \s_1 \psi (-x) \label{P-S}\\
P_L: && \psi (x) \rightarrow \s_2 \psi (-x) \label{P-L}
\eea
Using bosonization rules for spin-1/2 Dirac fermions (see e.g. \cite{GNT}),
one easily finds that both $P_S$ and $P_L$ lead to
$$
\Phi_c (x) \rightarrow - \Phi_c (-x),
$$
whereas the WZNW field $\hat{g}_s$ transforms differently:
\bea
P_S: && \hat{g}_s (x) \rightarrow - \hat{g}_s (- x) \label{g-PS}\\
P_L: && \hat{g}_s (x) \rightarrow \hat{g}_s (- x)\label{g-PL}
\eea
Comparing (\ref{delta-term}) and (\ref{dimer-op-den}), we see that
the $\Delta$-perturbation breaks $P_L$ but is invariant under $P_S$
(in fact, $P_S$ is the symmetry of the Hamiltonian (\ref{ham})), while
the dimerization operator $D$ breaks $P_S$ but preserves $P_L$.
This is easily understood by noticing that these two terms have the structure
of two different fermionic mass bilinears,
$ \psi^{\dagger} \s_1 \psi$ and $ \psi^{\dagger} \s_2 \psi$,
with opposite transformation properties with respect to $P_S$ and $P_L$.
Thus, the parity properties of the SDI and BI phases are different,
and this is a strong indication that the passage from SDI to BI should
be associated with a significant redistribution of the charge density
(charge transition).

In what follows, we will not be dealing with
estimation of the transition point $U_{c1}$.
Referring the reader to Ref.\cite{FGN} where the mechanism of the charge
transition is discussed in the context of excitonic instability of the BI
phase, here we simply claim that for the model (\ref{ham}), within
the leading logarithmic accuracy, 
$U_{c2}/U_{c1} -1 = {\em const}/\ln \left(t/\Delta \right)$,
where the positive constant is of the order of unity.
Let us instead focus on a nonperturbative description of the
charge transition to the BI phase. Suppose we are in the SDI phase with
both charge and spin sectors gaped. 
Adopting an Abelian bosonic representation for the $SU(2)_1$ WZNW model with 
a marginally {\sl relevant} perturbation, 
\bea
{\cal H}_s &=& \frac{v_s}{2} \left[ \Pi^2 _s + \left(\p_x \Phi_s  \right)^2 
\right]
\nonumber\\
&-& \frac{\lambda}{\pi} \p_x \Phi_{sR} \p_x \Phi_{sL}
+ \frac{\lambda}{2 (\pi \alpha)^2} \cos \sqrt{8\pi} \Phi_s, 
\label{spin-Abel}
\eea
we can regard the effective Hamiltonian given by (\ref{spin-Abel}),
(\ref{charge}) and (\ref{delta-term}) as a phenomenological Landau-Ginzburg
energy functional, in the sense that all the couplings ($m_0, K_c, \lambda, 
\Delta$)
and velocities ($v_s, v_c$) are effective ones obtained by
integrating out high-energy degrees of freedom.
The effective potential is given by:
\be
{\cal U} (\Phi_c , \Phi_s) = - \mu_c \cos \sqrt{8\pi K_c}\Phi_c - 
\mu_s \cos \sqrt{8\pi}\Phi_s
- \delta \sin \sqrt{2\pi K_c}\Phi_c \cos \sqrt{2\pi}\Phi_s
\label{U-eff} 
\ee
with
$$
\mu_c = \frac{m_0}{\pi \alpha} > 0,
~~~~\mu_s = \frac{\lambda}{2 (\pi \alpha)^2} > 0,~~~~
\delta = \frac{\Delta}{\pi \alpha}.
$$
The robust ingredients of the potential (\ref{U-eff}) are
the vertex operators and the signs of the corresponding amplitudes.
It can be shown that, as long as the parameter $K_c$ 
is confined within the interval $1/2 < K_c < 1$,
this potential, with all terms being strongly relevant perturbations, 
is indeed the most representative
one for the model under discussion because no new relevant operators are 
generated
in the course of renormalization. This is no longer true if $K_c < 1/2$, the 
regime
which can be realized for an extended model with extra finite-range 
interactions,
e.g. $V \sum_i n_i n_{i+1}$. In this case, new relevant vertex operators will
be generated upon renormalization, and the continuous Ising transition 
in the charge sector can
transform to a first-order one. We will comment on that in the end of this 
section.

A simple analysis of the saddle points of the potential 
${\cal U} (\var_c , \var_s)$\cite{FGN}
shows that the location of its minima in the spin sector, 
$\var_s = \sqrt{\pi/2}n$,
and hence the spin quantum
numbers of the topological excitations, are
the same as in the BI phase ($U < U_{c1}$).
So the spin part of the spectrum in the SDI phase smoothly transforms
to that of the BI phase.
Therefore, being interested in the redistribution of the charge degrees of 
freedom
in the vicinity of $U_{c1}$, in (\ref{U-eff}) we can replace 
$\cos \sqrt{2\pi} \Phi_s$ by its vacuum expectation value,
$\la \cos \sqrt{2\pi} \Phi_s \ra = \pm c_0$. The two signs here
reflect the
$Z_2$ degeneracy of the dimerized ground state (spontaneously broken
site parity; see Eq. (\ref{g-PS})).
Thus, the Hamiltonian of effective model describing the charge degrees of 
freedom 
can be represented in the following 2$\times$2 matrix form:
\be
H_{\em c;eff} = \left(
\begin{array}{clcr}
H^{(+)}_c&&0\\
0&&H^{(-)}_c
\end{array}
\right)
\label{H-eff-matrix}
\ee
where
\bea
H^{(\pm)}_c &=& \frac{v_c}{2} \left[ \left( \p_x \Phi_c \right)^2 
+ \left( \p_x \Theta_c \right)^2 \right] \nonumber\\
&-& \mu_c \cos \sqrt{8\pi K_c} \Phi_c \mp h \sin \sqrt{2\pi K_c} \Phi_c
\label{charge-ham}
\eea
(here $h = \delta c_0$). Notice that under $P_S$ $~H^{(+)}_c \leftrightarrow
H^{(-)}_c$.

We have arrived at the matrix version of DSG model similar to
(\ref{matrixDGS}), (\ref{QAT-DSG}). To 
make contact with the DQAT model discussed in detail in sections 2 and 3,
we shall consider $H_{\em c;eff}$ in the vicinity of the point $K_c = 1/2$.
Setting $K_c = 1/2 \left(1 + \gamma_0 \right)$ and rescaling the fields
$$
\Phi_c = \frac{1}{\sqrt{2K_c}}~ \Phi, ~~~
\Theta_c = \sqrt{2K_c}~ \Theta
$$
transforms $H^{(\pm)}_c$ in (\ref{charge-ham}) to a form similar 
to ${\cal H}_{\pm}$ of Eq.(\ref{QAT-DSG}).

With this correspondence and all the results of the previous sections, we are 
now
able to describe in detail
the Ising transition in the charge sector. As already explained in section 2.2,
this can be done by considering only the Hamiltonian $H^{(+)}_c$.

When studying the physical properties of our model (\ref{ham}) at the charge
transition point, it is important to remember that the disordered ($h < h_c$)
and ordered ($h > h_c$) phases of the effective Ising model
correspond to the SDI and BI phases of the electronic model (\ref{ham}).
First we consider the dimerization operator (\ref{dimer-op-den}):
$$
{\cal D} \sim \cos \sqrt{2\pi K_c} \Phi_c \cos \sqrt{2\pi} \Phi_s.
$$
With the spin field $\Phi_s$ locked in the SDI phase,
$\la \cos \sqrt{2\pi} \Phi_s \ra = +c_0$ and $K_c =1/2$, this transforms to
the charge polarisation field
$$
{\cal D} \sim \cos \sqrt{\pi} \Phi_c.
$$
In full agreement with the physical picture, from (\ref{cos}) it follows that
the operator ${\cal D}$, being the 
order parameter of the SDI phase, is the most strongly fluctuating field
at the Ising trasntion:
\be
{\cal D} \sim \mu \label{D-mu}
\ee
Its average value is nonzero in the SDI phase and vanishes
as $\la {\cal D} \ra \sim (h_c - h)^{1/8}$ on approaching the charge transition
point (remaining zero in the whole BI phase).

Using (\ref{sin}), we find that
the average value of the $\Delta$-perturbation, which, at $K_c = 1/2$, is
given by
$
{\cal O}_{\Delta} \sim \sin {\pi} \Phi_c 
$
is nonsingular across the transition and remains finite in both
phases.

The most interesting feature of the Ising transition 
is the UV-IR transmutation of the charge density $\rho_c (x)$
and current $J_c (x)$. At $K_c = 1/2$, the UV limit of our model 
represents a metallic state with central charge $c_{UV} = 2$,
described in terms of two massless Gaussian fields $\Phi_c$ and $\Phi_s$.
In this limit
$$
\rho_c (x) = \frac{1}{\sqrt{\pi}} \p_x \Phi_c (x), ~~~
J_c (x) = - \frac{1}{\sqrt{\pi}} \p_x \Theta_c (x).
$$
According to (\ref{csum}),(\ref{c5sum}), at the Ising criticality
($c_{IR} = 1/2$),
\be
\rho_c (x) = C \p_x \mu (x), ~~~J_c (x) = - C \p_t \mu (x) 
\label{rho-J-IR}
\ee
With $\mu (x)$ representing the charge polarization field,
Eqs.(\ref{rho-J-IR}) identify $\rho_c$ and $J_c$ as the bound charge
and polarisation-current densities, 
respectively. Such an identification is typical
for insulators rather than metals. 
This could have been anticipated from the fact that a true metallic state
with charge-carrying gapless excitations cannot be described by a single
massless real (Majorana) field.

An insulating (semi-metallic) behaviour of the model at the 
quantum critical point, reached at the
charge transition,
becomes manifest when one estimates the optical conductivity:
\be
\s (\omega) \sim - \omega \Im m D^{(R)}(\omega, 0).
\label{optcond-def}
\ee
Here $D^{(R)}(\omega, q)$ is the retarded correlation function 
defined in (\ref{ret}). Using the result (\ref{Dnotfin}), we find that,
at zero temperature, $\s(\omega)$ displays a universal power-law
behaviour:
\be
\s_0 (\omega) \sim \omega^{-3/4} \label{optcond-T=0}
\ee
Although the optical conductivity is divergent in the zero-frequency limit,
there is no Drude-peak contribution ($\sim \delta(\omega)$) typical
of true metals.

It is also possible to estimate the optical conductivity at finite temperatures
(keeping in mind the $\s-\tau$ representation of the DQAT model, one
should assume that $T$ is much smaller than
the mass gap in the decoupled $\tau$ degrees of freedom). This can be done
using conformal mapping from
a cylinder $[-\infty < x < \infty; ~0 < \tau < \beta ]$ onto a complex
plane $C$. Starting with the $T=0$ asymptotics of the correlation function
$$
\la \mu (z_1, \bar{z}_1) \mu (z_2, \bar{z}_2) \ra
\propto \frac{1}{|z_1 - z_2|^{1/4}},
$$
and using the above mentioned conformal mapping, one obtains
\be
\la \mu (x,\tau) \mu (0,0) \ra \propto
\left[ \frac{\pi T}{\sinh \pi T(x - \ri \tau)
\sinh \pi T(x + \ri \tau)} \right]^{1/8}.
\label{mu-mu-corr-finite-T}
\ee
It can be shown that\cite{SB}
\be
- \Im m \chi(\omega) \sim \frac{1}{T^{7/4}} \Im m ~
\left [ \rho \left(\frac{\omega}{4\pi T}  \right) \right]^2,
\label{im-chi}
\ee
where
\be
\rho(x) = \frac{\Gamma \left(\frac{1}{16} - \ri x \right)}
{\Gamma \left(\frac{15}{16} - \ri x \right)}.
\label{ro}
\ee
The final result for the temperature dependent 
optical conductivity at the Ising
critical point is given by the following {\sl universal} formula:
\be
\s (\omega, T) \propto \frac{\omega}{T^{7/4}}
\Im m ~
\left [ \rho \left(\frac{\omega}{4\pi T}  \right) \right]^2
\label{sigma-at-finite-T}
\ee

At finite $\omega \ll T$, the frequency dependence of $\s (\omega, T)$
is classical but with a quantum, temperature dependent prefactor:
$$
\s (\omega, T) \sim \omega^2 / T^{11/4}.
$$
At $\omega \sim T$, $\s (\omega, T)$ reaches its maximum and then crosses over
to its quantum-critical high-frequency ($\omega \gg T$) asymptotics
(\ref{optcond-T=0}).

\section{First order transition at $\beta^2<2\pi$}
 
Until now we assumed that $2\pi <\beta^2<8\pi$, so that no 
other relevant operators were generated upon renormalization.
Indeed, already at $\beta^2<32\pi/9$, the operator 
$\sin\left[\left(3\beta/2\right)\Phi(x)\right]$, which is 
generated by the OPE of the two operators present in 
(\ref{HDSG}), becomes relevant. However, such an operator 
does not modify the qualitative behaviour of the model, as 
one can easily realize by a quasi-classical analysis. 
On the contrary, the operator $\cos\left(2\beta\Phi(x)\right)$, 
which is also generated by the OPE, and which becomes 
relevant at $\beta^2<2\pi$, can modify the 
properties of the model in a relevant manner. Indeed,  
inspecting the quasi-classical potential
\[
{\cal U}\left[\Phi\right] = - g\,\cos\beta\Phi - 
\lambda\, \sin\left(\frac{\beta}{2}\Phi\right) 
- V\, \cos 2\beta\Phi,
\]
one finds that the Ising transition is turned by a 
sufficiently large $V$ into a first order one.

The capability of an apparently subleading operator to change a 
continuous transition to a first order one is indeed common to a variety 
of models. For instance, it is known that the critical 
line with non universal exponents separating  
the Charge Density Wave (CDW) and the Spin Density Wave (SDW) phases of the 
extended (U-V) Hubbard model, becomes a first order line at sufficiently 
strong coupling\cite{Hirsh}. In fact, in the extended Hubbard model,  
the charge Luttinger liquid exponent $K_c$ can be lower than $1/2$, 
the point at which the second harmonics of the Umklapp scattering 
starts to be relevant. Therefore, at sufficiently strong interaction, it 
can indeed turn the CDW--SDW transition line into a first order one.
A similar situation occurs with the charge transition in the 
electronic model, considered in section 7, when the latter is generalised
to include a sufficiently strong nearest-neighbour repulsion \cite{nagaosa}.   

\section{Conclusions}

In this paper, we have proposed a nonperturbative description of the
Ising criticality in the DSG model (\ref{HDSG}). 
Using the equivalence between the DSG model and a deformed 
quantum Ashkin-Teller model, valid in the vicinity of
the decoupling point, $\beta^2 = 4\pi$,
we were able to identify
the effective Ising degrees of freedom that asymptotically
decouple from the rest of the spectrum and become critical 
in the infrared limit. This identification allowed us to describe
the UV-IR ``transmutation'' of all physical fields of the DSG model
and calculate the correlation functions at and close to the transition.
We have also demonstrated the efficiency of our approach to
describe Ising transitions in some physical realizations of the
DSG model.

We believe that our quantum-Ising-chain approach can be generalised 
to the case when the number of the constituent Ising models,
coupled by the interaction $h \prod_j \s_j$, is larger than 2.
Such situation can indeed be realized in certain SU(2)-invariant
spin-ladders models which can be driven to criticality 
under the action of external staggered fields. 
In such systems, the quantum critical points may be not only of
the Ising type but also correspond
to SU(2)$_k$ WZNW universality class.
Such examples will be considered elsewhere \cite{{else}}.  
\bigskip\bigskip

{\bf Acknowledgements} 

It is our pleasure to thank G. Mussardo for inspiring discussions.
We are grateful A. Chubukov, D. Edwards, Yu Lu, N. Nagaosa, 
S. Sorella, E. Tosatti,
A. M. Tsvelik, Y.-J. Wang and V. Yakovenko for their
interest in the work and helpful comments.
M.F. is partly supported by INFM, under project PRA HTSC.
A. O. G. is supported the EPSRC of the United Kingdom.
A. A. N. is partly supported by INTAS-Georgia grant 97-1340.

\newpage
\appendix

\section{Bosonization}
\subsection{Bosonization of Fermi fields, currents and mass bilinears}

We build up a Dirac field out of two Majorana fields, $\xi_1$ and
$\xi_2$, and bosonize it:
\be
\psi = \left(
\begin{array}{clcr}
R\\
L
\end{array}
\right)
= \left[  \frac{\xi_1 + \ri \xi_2}{\sqrt{2}} \right]_{R,L} \Rightarrow
\frac{1}{\sqrt{2\pi \alpha}} 
\exp \left( \pm \ri \sqrt{4\pi}\phi_{R,L} \right)
\label{Maj-Dirac-Bose}
\ee
To ensure anticommutation between the right and left components of the Fermi 
field,
it is assumed that
\be
\left[\phi_R , \phi_L  \right] = \frac{\ri}{4}.\label{RL-comm}
\ee
Notice that $\alpha$ is an ultraviolet cutoff in the {\sl bosonic} theory which
actually appears as a short-distance regulator in the normal-mode expansion of
bosonic fields. For this reason it does not need to coincide with the original 
lattice
constant $a_0$ which appears in the continuum representation of {\sl fermionic}
fields. These two cutoffs are, however, related, as shown in Appendix A.2.

The chiral components of the U(1) current
are defined as
\bea
J_{R} &=& :R^{\dagger} R: = \ri \xi_{1R} \xi_{2R}
= \frac{1}{\sqrt{\pi}} \p_x \phi_R \label{cur-R}\\
J_L &=& :L^{\dagger}L: = \ri \xi_{1L} \xi_{2L}
= \frac{1}{\sqrt{\pi}} \p_x \phi_L
\label{cur-L}
\eea
Using (\ref{Maj-Dirac-Bose}), one also finds that
\bea
R^{\dagger}L &=& - \frac{\ri}{2\pi \alpha} e^{- \ri \sqrt{4\pi} \Phi}
\nonumber\\
&=& \frac{1}{2} \left(\xi_{1R} - \ri \xi_{2R}  \right)
\left(\xi_{1L} + \ri \xi_{2L}  \right), \label{R^+-L}
\eea
implying that
\bea
\cos \sqrt{4\pi} \Phi &=& \ri \pi \alpha
\left(\xi_{1R} \xi_{1L} + \xi_{2R} \xi_{2L} \right) \label{cos-Maj}\\
\sin \sqrt{4\pi} \Phi &=& - \ri \pi \alpha
\left(\xi_{1R} \xi_{2L} + \xi_{1L} \xi_{2R} \right)
\label{sin-Maj}
\eea

The DQAT Hamiltonian (\ref{DQAT}) is symmetric under interchange
$({\cal P}_{12})$ of the two chains. Let us find the transformation properties 
of
all the fields under ${\cal P}_{12}$.
>From (\ref{Maj-Dirac-Bose}) it follows that interchanging the two chains 
leads to transformations
\be
\phi_R \rightarrow \frac{\sqrt{\pi}}{4} - \phi_R, ~~~
\phi_L \rightarrow - \frac{\sqrt{\pi}}{4} - \phi_L
\label{P-12-phi-RL}
\ee
The chiral currents $J_{R,L}$, and therefore the total $(J = J_{R} + J_{L})$
and axial  $(J_5 = J_{R} - J_{L})$ currents change their signs:
\be
J_{R(L)} \rightarrow - J_{R(L)}, ~~
J  \rightarrow - J, ~~J_5  \rightarrow - J_5
\label{P-12-curr}
\ee
Under ${\cal P}_{12}$ the scalar field $\Phi = \phi_R + \phi_L$ and its dual 
counterpart
$\Theta = - \phi_R + \phi_L$ transform as follows:
 \be
\Phi \rightarrow - \Phi, ~~~\Theta \rightarrow - \frac{\sqrt{\pi}}{2} - \Theta
\label{P-12-Phi-Theta}
\ee
Notice that the symmetry properties of r.h.sides of (\ref{cos-Maj})and
(\ref{sin-Maj}) are consistent with (\ref{P-12-Phi-Theta}).

\subsection{Bosonization of products of order and disorder operators}

Starting with the Jordan--Wigner transformation for two QI spin chains,
which can be summarised as follows 
\bea
\s^x _{an} = \ri \zeta _{an}  \eta_{an}, &&
\mu^x _{a,n+1/2} = - \ri \zeta_{an} \eta_{a,n+1} \label{a1}\\
\s^z _{an} = \ri \kappa_a \left( \prod_{j=1}^{n} \s^x _{aj} \right) \zeta_{an}, 
&&
\mu^z _{a,n+1/2} =  \prod_{j=1}^{n}\s^x _{aj} \label{b1}\\
\s^y _{an} =  \ri \kappa_a \left( \prod_{j=1}^{n} \s^x _{aj} \right)\eta_{an}, 
&& 
\mu^y _{a,n+1/2} =  \left( \prod_{j=1}^{n}\s^x _{aj} \right),
\eta_{a,n+1} \zeta_{an} \label{c1} 
\eea
here we derive bosonized expressions for products of Ising fields belonging to
different chains.

\medskip

$\bullet~$ \underline{$\mu^z _{1,n+1/2}~ \mu^z _{2,n+1/2}$}$~$.
Using the lattice definition (\ref{b1}), we have:
\be
\mu^z _{1,n+1/2} \mu^z _{2,n+1/2}
= \prod_{j=1}^{n} \s^x _{1j} \s^x _{2j}
\label{mumu-1}
\ee
There are many ways to exponentiate the product in (\ref{mumu-1}), the naive 
assumption
would be to use the identity
$$
\s^x = \mp \ri e^{\pm \ri (\pi/2) \s^x}
$$
and then replace in the exponential $\s^x$ by $\ri \eta \zeta$. This will bring 
us
to a phase proportional to
$$
\sum_{j=1}^{n} \left( \eta_{1n} \zeta_{1n} +\eta_{2n} \zeta_{2n} \right)  
$$
which, in the continuum limit, reduces to
$$
\int_{0}^{x} \rd x' \left( \xi_{1R} \xi_{1L} + \xi_{2R} \xi_{2L} \right).
$$
According to (\ref{cos-Maj}), the integrand represents a bosonic cosine 
operator,
which is not what we would like to have for practical purposes.
To get a better representation, one has first to rearrange the four Majorana
fields in
the product $\s^x _{1j} \s^x _{2j}$:
\be
\s^x _{1j} \s^x _{2j} 
= \left( \eta_{1j} \eta_{2j} \right) \left(\zeta_{1j} \zeta_{2j} \right) 
\ee
It is readily seen that  a product of two local Majorana fields, defined on the 
lattice,
can be represented as
\be
\eta_1 \eta_2 = \pm \exp \left( \pm \frac{\pi}{2}\eta_1 \eta_2  \right)
\ee
Therefore
\be
\mu^z _{1,n+1/2} \mu^z _{2,n+1/2} = \prod_{j=1}^n \s^x _{1j} \s^x _{2j} =
\exp \left[ \pm \frac{\pi}{2}
\sum_{j=1}^n \left( \eta_{1j} \eta_{2j} + \zeta_{1j} \zeta_{2j}\right)
 \right]
\ee

Now we pass to the continuum limit by making chiral rotation (\ref{chiral-rot2}) 
and using (\ref{cur-R}), (\ref{cur-L}):
\bea
\mu^z _{1,n+1/2} \mu^z _{2,n+1/2} &\rightarrow&
\exp \left[\pm \pi \int_0 ^x \rd y~\left( \xi_{1R} \xi_{2R}
+ \xi_{1L} \xi_{2L}  \right) \right]\nonumber\\
&=& \exp \left[ \mp \ri \sqrt{\pi} \int_0 ^x \rd y~ \p_y \Phi (y)
\right] \nonumber\\
&=& e^{\mp \ri \sqrt{\pi} \Phi (x)} \label{mumu-intermed1}
\eea
The l.h.s. of this relation is symmetric under ${\cal P}_{12}$, so must
the r.h.side. As follows from (\ref{P-12-Phi-Theta}),
under ${\cal P}_{12}$ the field $\Phi$ changes its sign.
So the sign ambiguity in (\ref{mumu-intermed1}) is resolved by
replacing the phase exponential by a cosine:
\be
\mu^z _{1,n+1/2} \mu^z _{2,n+1/2} =  \cos \sqrt{\pi} \Phi(x)
\label{mumu-fin}
\ee
\medskip

$\bullet~$ \underline{$\s^z _{1,n+1/2}~ \s^z _{2,n+1/2}$}.$~$
Using (\ref{b1}), we have:
\bea
\s^z _{1n} \s^z _{2n} &=& - \left( \kappa_1 \kappa_2 \right)
\left( \zeta_{1n}\zeta_{2n}\right)
\left( \prod_{j=1}^{n} \s^x _{1j} \s^x _{2j} \right)
 \nonumber\\
&=& - \kappa_1 \kappa_2~ \zeta_{1n}\zeta_{2n} 
\left(\mu^z _{1,n+1/2} \mu^z _{2,n+1/2} \right)
\label{ss-1}
\eea
With the representation (\ref{mumu-fin}) at hand,
we only need to pass to the continuum limit in $\zeta_{1n}\zeta_{2n}$
and then make use of proper operator product expansions. We have:
\bea
\zeta_{1n}\zeta_{2n} &\rightarrow& 2a_0 \zeta_1 (x) \zeta_2 (x) \nonumber\\
&=& a_0 \left[ \xi_{1R} (x) + \xi_{1L} (x) \right]
\left[ \xi_{2R} (x) + \xi_{2L} (x) \right]\nonumber\\
&=& - \frac{\ri a_0}{\sqrt{\pi}} \p_x \Phi (x) + \frac{\ri a_0}{\pi \alpha}
\sin \sqrt{4\pi} \Phi(x)
\label{zeta-zeta-cont}
\eea
So
\be
\s^z _{1n} \s^z _{2n} \rightarrow \ri \kappa_1 \kappa_2 a_0
\left[ \frac{1}{\sqrt{\pi}} \p_x \Phi (x) - \frac{1}{\pi \alpha}
\sin \sqrt{4\pi} \Phi(x) \right] \cos \sqrt{\pi}\Phi (x + \alpha )
\ee

Picking up the most relevant operators in the following OPE
\bea
\p_x \Phi (x) :\cos \sqrt{\pi} \Phi (x \mp \alpha):
&=& \pm \frac{1}{2\sqrt{\pi} \alpha} :\sin \sqrt{\pi} \Phi (x):
+ \cdots
\label{ope-1}\\
:\sin \sqrt{4 \pi} \Phi (x)::\cos \sqrt{\pi} \Phi (x):
&=& \frac{1}{2} :\sin \sqrt{\pi} \Phi (x): + \cdots
\label{ope-2}
\eea
one finds that
\be
\s^z _{1n} \s^z _{2n} \rightarrow
- \ri \kappa_1 \kappa_2 \left( \frac{a_0}{2\pi \alpha} \right)
:\sin \sqrt{\pi} \Phi (x):
\label{ss-int}
\ee
Intending to keep duality among (\ref{ss-int}) and (\ref{mumu-fin})
we set
\be
\alpha = \frac{a_0}{\pi} \label{alpha-vs-a0}
\ee
and finally obtain:
\be
\s^z _{1n} \s^z _{2n} =
- \ri \kappa_1 \kappa_2 :\sin \sqrt{\pi} \Phi (x):
\label{ss-fin}
\ee

Notice the important role of the Klein product $\kappa_1 \kappa_2$.
The l.h.s. of (\ref{ss-fin}) is ${\cal P}_{12}$-symmetric while
in the r.h.side $\sin \sqrt{\pi} \Phi (x)$ is antisymmetric 
(see (\ref{P-12-Phi-Theta})).
So the r.h.s. of (\ref{ss-fin}) is symmetric just due to
the presence of $\kappa_1 \kappa_2$. If we had replaced $\kappa_1 \kappa_2$
by a constant, $\kappa_1 \kappa_2 \rightarrow \ri$,
we would obtain
$$
\s^z _{1n} \s^z _{2n} = :\sin \sqrt{\pi} \Phi (x):
$$
In this representation, the only way
to ensure the ${\cal P}_{12}$-symmetry would be to 
impose a constraint that identifies $\Phi$ and $- \Phi$, in which case
the scalar field $\Phi$ would transform to an orbifold.

\medskip
$\bullet~$ \underline{$\s^z _{1n} \mu^z _{2,n+1/2}$}.$~$
First we represent $\s^z _{1n} \mu^z _{2,n+1/2}$ as
$$
\s^z _{1n} \mu^z _{2,n+1/2} = 
\left( \s^z _{1n} \mu^z _{1,n+1/2} \right) 
\left( \mu^z _{1,n+1/2}  \mu^z _{2,n+1/2}\right).
$$
The continuum limit for second product has already 
been found (see Eq.(\ref{mumu-fin})).
The first product reduces to  a Majorana field:
$
\s^z _{1n} \mu^z _{1,n+1/2} = - \ri \kappa_1 \zeta_{1n}.
$
In the continuum limit, with relation (\ref{alpha-vs-a0}) taken into account,
\bea
\zeta_{1n} &\rightarrow& \sqrt{a_0} \left[ \xi_{1R} (x) +
\xi_{1L} (x) \right]\nonumber\\
&=& \cos \sqrt{4\pi} \Phi_R (x) + \cos \sqrt{4\pi} \Phi_L (x) 
\label{zeta_1-cont}
\eea
So
\be
\s^z _{1n} \mu^z _{2,n+1/2} = - \ri \kappa_1 
\left[ \cos \sqrt{4\pi} \Phi_R (x) + \cos \sqrt{4\pi} \Phi_L (x)  \right]
\cos \sqrt{\pi} \Phi (x + \alpha)
\label{smu-int1}
\ee
Making use of the following OPE
\bea
:\cos \sqrt{4\pi} \Phi_R (x)::\cos \sqrt{\pi} \Phi (x + \alpha):
&=& :\cos \sqrt{4\pi} \Phi_L (x)::\cos \sqrt{\pi} \Phi (x + \alpha):\nonumber\\
&=& \frac{1}{2} :\cos \sqrt{\pi} \Theta(x): + \cdots 
\label{ope-int-e}
\eea
we arrive at the result:
\be
\s^z _{1n} \mu^z _{2,n+1/2} =
- \ri \kappa_1 :\cos \sqrt{\pi} \Theta(x):
\label{smu-fin}
\ee

\medskip
$\bullet~$ \underline{$\mu^z _{1,n+1/2}\s^z _{2n}$}.$~$
Quite similarly one finds that
\be
\mu^z _{1,n+1/2} \s^z _{2n}=
\ri \kappa_2 :\sin \sqrt{\pi} \Theta(x):
\label{mus-fin}
\ee
Using (\ref{P-12-Phi-Theta}), we see 
that under ${\cal P}_{12}$ the r.h.sides of Eqs.(\ref{smu-fin}) and 
(\ref{mus-fin})
indeed transform to each other.

\newpage

\section{More on lattice fermions}
\label{Lattice-fermions}

An alternative way to bosonize the QAT model, is to pass through a 
mapping onto spinless fermions, and bosonize the latter. 

We start by the quantum Ising model (\ref{QIM}). By 
means of the following unitary transformation
\bea
\zeta_n &=&-\frac{1}{\sqrt{2}}\left(\xi_{Rn}-\xi_{Ln}\right),\\ 
\eta_n &=&\frac{1}{\sqrt{2}}\left(\xi_{Rn}+\xi_{Ln}\right),
\eea
the Hamiltonian (\ref{QIM}) transforms onto
\bea
H_{QI} &=& \frac{J}{2}\left(-\frac{i}{2}\right)\sum_n
\left[ 
\xi_{Rn}\left(\xi_{Rn+1}-\xi_{Rn-1}\right)
-\xi_{Ln}\left(\xi_{Ln+1}-\xi_{Ln-1}\right) \right] \nonumber\\
&+&i\frac{J}{4}\sum_n \left[
\xi_{Rn}\left(2\xi_{Ln}-\xi_{Ln+1}-\xi_{Ln-1}\right)
-\xi_{Ln}\left(2\xi_{Rn}-\xi_{Rn+1}-\xi_{Rn-1}\right)\right]\nonumber\\
&+& i\left(\Delta_J-J\right)\sum_n \xi_{Rn}\xi_{Ln}.
\label{HIsing}
\eea

We now consider the following quantum Ashkin--Teller Hamiltonian 
of two-coupled Ising chains
\be
H_{QAT} = H_{QI}\left[\sigma_1\right] + H_{QI}\left[\sigma_2\right] 
+H'_{AT}\left[\sigma_1,\sigma_2\right],
\label{HAskin}
\ee
where $H_{QI}$'s are Ising Hamiltonians [see Eq.(\ref{HIsing})] 
for two spin species, $\sigma_1$ and $\sigma_2$, and 
\be
H'_{AT} = K\sum_n \left(
\sigma^z_{1,n}\sigma^z_{1,n+1}\sigma^z_{2,n}\sigma^z_{2,n+1} 
+\sigma^x_{1,n}\sigma^x_{2,n}\right),
\label{HAT}
\ee
is the coupling term. 
For each spin species we introduce Majorana's fermions. In 
order to make $\sigma_1$ and $\sigma_2$ commute we multiply 
each Majorana's fermion for the chain 1 by another Majorana 
$\kappa_1$, and for chain 2 by $\kappa_2$. That is, if $a=1,2$, 
we define
\be
\zeta_{a,n} = \kappa_a \left(i\sigma^z_{a,n}\mu^z_{a,n+\frac{1}{2}}\right)
,\;\;
\eta_{a,n} = \kappa_a \left(\sigma^z_{a,n}\mu^z_{a,n-\frac{1}{2}}\right),
\label{2SM}
\ee
as well as their $R,L$ components
\bea
\zeta_{1,n} &=&-\frac{1}{\sqrt{2}}\left(\xi_{Rn}-\xi_{Ln}\right),\\ 
\eta_{1,n} &=&\frac{1}{\sqrt{2}}\left(\xi_{Rn}+\xi_{Ln}\right),\\
\zeta_{2,n} &=&-\frac{1}{\sqrt{2}}\left(\eta_{Rn}-\eta_{Ln}\right),\\ 
\eta_{2,n} &=&\frac{1}{\sqrt{2}}\left(\eta_{Rn}+\eta_{Ln}\right).
\eea
Let us concentrate for the moment onto the two Ising Hamiltonians, which 
are bilinear in the Majorana fermions. We notice the following 
general property ($p,q=R,L$)
\bn
&&\xi_{pn}\xi_{qm}+\eta_{pn}\eta_{qm} = 
\frac{1}{2}\left[ 
\left(\xi_{pn}-i\eta_{pn}\right)
\left(\xi_{qm}+i\eta_{qm}\right)
+ \left(\xi_{pn}+i\eta_{pn}\right)
\left(\xi_{qm}-i\eta_{qm}\right)\right] \\
&=& 
2\left[ c^\dagger_{pn}c^\pdag_{qm} + c^\pdag_{pn}c^\dagger_{qm}\right]
=2\left[ c^\dagger_{pn}c^\pdag_{qm} - c^\dagger_{qm}c^\pdag_{pn} 
+ \delta_{pq}\delta_{nm}\right],
\en
where we define the Fermi operators
\be
c_{R(L)n} = \frac{1}{2}\left( \xi_{R(L)n}+i\eta_{R(L)n}\right). 
\label{def:fermion}
\ee
Therefore, through (\ref{HIsing}), we find
\bea
&&H_{QI}\left[\sigma_1\right] + H_{QI}\left[\sigma_2\right] = \nonumber\\ 
&=&J\left(-\frac{i}{2}\right)\sum_n
\left[ 
c^\dagger_{Rn}\left(c^\pdag_{Rn+1}-c^\pdag_{Rn-1}\right)
-c^\dagger_{Ln}\left(c^\pdag_{Ln+1}-c^\pdag_{Ln-1}\right) - H.c.\right] 
\nonumber\\
&&+i\frac{J}{2}\sum_n \left[
c^\dagger_{Rn}\left(2c^\pdag_{Ln}-c^\pdag_{Ln+1}-c^\pdag_{Ln-1}\right)
-c^\dagger_{Ln}\left(2c^\pdag_{Rn}-c^\pdag_{Rn+1}-c^\pdag_{Rn-1}\right)
- H.c.\right]
\nonumber\\
&& +2i\left(\Delta_J-J\right)\sum_n c^\dagger_{Rn}c^\pdag_{Ln} 
-c^\dagger_{Ln}c^\pdag_{Rn}.
\label{2Ising}
\eea

The analogy with a fermionic model on a lattice is already apparent. 
Let us make this analogy more firm. We consider the following 
spinless fermion model on a lattice of $2L$ sites 
\bn 
H &=& -t\sum_{n=0}^{2L-1} c^\dagger_n c^\pdag_{n+1} + H.c.\\
&=& -t\sum_{n=0}^{L-1} c^\dagger_{2n+1}\left(c^\pdag_{2n} 
+ c^\pdag_{2n+2}\right) 
+c^\dagger_{2n}\left(c^\pdag_{2n-1}+c^\pdag_{2n+1}\right).
\en
We make the following unitary transformation
\bn
c_{2n} &=& \frac{(-1)^n}{\sqrt{2}}\left(c_{Rn}+c_{Ln}\right),\\
c_{2n+1} &=& i\frac{(-1)^n}{\sqrt{2}}\left(c_{Rn}-c_{Ln}\right),
\en
where the right- and left-moving fermions are defined on a lattice of 
half the number of sites, i.e. $L$. The Hamiltonian becomes
\bea
H&=& -i\frac{t}{2}\sum_n  
c^\dagger_{Rn}\left(c^\pdag_{Rn+1}-c^\pdag_{Rn-1}\right)
-c^\dagger_{Ln}\left(c^\pdag_{Ln+1}-c^\pdag_{Ln-1}\right)\nonumber\\
&+&i\frac{t}{2}\sum_n 
c^\dagger_{Rn}\left(2c^\pdag_{Ln}-c^\pdag_{Ln+1}-c^\pdag_{Ln-1}\right)
-c^\dagger_{Ln}\left(2c^\pdag_{Rn}-c^\pdag_{Rn+1}-c^\pdag_{Rn-1}\right).
\nonumber\\
\label{Hsf}
\eea
Let us add to this Hamiltonian a staggered hopping
\bea
\delta H &=& -\Delta\sum_n c^\dagger_{2n}c^\pdag_{2n+1} + H.c.\\
&=& i\Delta\sum_n c^\dagger_{Rn}c^\pdag_{Ln} - H.c.\, . 
\label{dHsf}
\eea
We notice that (\ref{Hsf}) plus 
(\ref{dHsf}) coincide with (\ref{2Ising}) if $2J = t$ and $2(\Delta_J-J)
=\Delta$. 

Now we focus on the coupling term (\ref{HAT}), which, in terms of 
Majorana's fermions reads
\[
H'_{AT} = K\sum_n \zeta_{1n}\zeta_{2n}\left(
\eta_{1n+1}\eta_{2n+1}+ \eta_{1n}\eta_{2n}\right).
\]
Transforming into right- and left-moving Majorana fermions we find that 
\bn
\zeta_{1n}\zeta_{2n} &=& \frac{1}{2}
\left(\xi_{Rn}-\xi_{Ln}\right)
\left(\eta_{Rn}-\eta_{Ln}\right)\\
&=&\frac{1}{2}\left(\xi_{Rn}\eta_{Rn}+\xi_{Ln}\eta_{Ln}
-\xi_{Rn}\eta_{Ln}-\xi_{Ln}\eta_{Rn}\right)\\
&=&-i\left(\rho_{Rn}+\rho_{Ln}-1-\Delta_n\right),
\en
where we used (\ref{def:fermion}), and we defined
$\rho_{pn}=c^\dagger_{pn}c^\pdag_{pn}$ ($p=R,L$) as well as  
$\Delta_n = c^\dagger_{Rn}c^\pdag_{Ln} + H.c.$. 
We can equivalently show that 
\[
\eta_{1n}\eta_{2n} = -i\left(\rho_{Rn}+\rho_{Ln}-1+\Delta_n\right).
\]
Therefore the coupling term can be written as 
\bea
H'_{AT} &=& - K\sum_n\left(\rho_{Rn}+\rho_{Ln}-1-\Delta_n\right)
\left(\rho_{Rn}+\rho_{Ln}-1+\Delta_n\right) \nonumber\\
&-& K\sum_n \left(\rho_{Rn}+\rho_{Ln}-1-\Delta_n\right)
\left(\rho_{Rn+1}+\rho_{Ln+1}-1+\Delta_{n+1}\right).
\label{Hint}
\eea
On the other hand let us consider the simplest nearest-neighbour 
interaction for spinless fermions, namely
\[
V\sum_n c^\dagger_{2n}c^\pdag_{2n} c^\dagger_{2n+1}c^\pdag_{2n+1}
+ c^\dagger_{2n+1}c^\pdag_{2n+1} c^\dagger_{2n+2}c^\pdag_{2n+2}.
\]
This interaction term turns out to be equivalent to (\ref{Hint}),  
apart from a chemical potential term, provided $4K=-V$. 
Notice that, if one considers an Ashkin-Teller coupling which is 
not self-dual, e.g.
\[
H'_{AT} = K_1\sum_n 
\sigma^z_{1,n}\sigma^z_{1,n+1}\sigma^z_{2,n}\sigma^z_{2,n+1} 
+ K_2\sum_n\sigma^x_{1,n}\sigma^x_{2,n},
\]
this translates into a staggered interaction for the spinless fermion 
model. This interaction, even without an explicit dimerization, is 
able to gap the fermionic spectrum, thus showing the importance of 
self-duality, even at the level of the 
coupling term (\ref{HAskin}), to get a critical behaviour.  

Therefore we have shown the equivalence between a model of two coupled 
Ising chains in a transverse field, given by the Hamiltonian  
(\ref{HAskin}), and a model of spinless fermion with nearest 
neighbour interaction and dimerized hopping. 

An interesting point in this respect regards quantum numbers. The spinless 
fermion model has for instance conserved number of particle. In $\rho$ 
is the density, in the reduced chain we must have that 
\be
\frac{1}{L}\sum_n \rho_{Rn}+\rho_{Ln} = 2\rho,
\label{conserved}
\ee
is conserved. On the other hand
\bn
&&\frac{1}{L}\sum_n (\rho_{Rn}+\rho_{Ln} -1) = 2\rho-1\\
&&\equiv -\frac{i}{2L}\sum_n \zeta_{1n}\zeta_{2n} + \eta_{1n}\eta_{2n},
\en
which, in terms of Ising variables, implies the conservation of a very 
non-local operator, which includes the spins and their dual counterparts. 

\subsection{Operator identities}
We can build up several operator identities in the lattice 
representation. 

\begin{itemize}
\item[(1)]The simplest one is obtained by the identity 
\be
2\left( \rho_{Rn}+\rho_{Ln} -1 \right) =   
i\left(
\zeta_{1n}\zeta_{2n} + \xi_{1n}\xi_{2n}\right),
\label{ID:1}
\ee
which relates the density operator of the spinless fermions to 
a particular bilinear of Majorana's fermions.

\item[(2)] The other identity derives from the equality
\begin{equation}
2\left(c^\dagger_{Rn}c^\pdag_{Ln}+ H.c.\right) =
i\left(\xi_{1n}\xi_{2n}-\zeta_{1n}\zeta_{2n}\right),
\label{B:CDW}
\end{equation}
which relates the charge density wave operator of the 
spinless fermions to the Majorana's fermions. 

\item[(3)] The last identity is obtained by the equality
\begin{equation}
-2i\left(c^\dagger_{Rn}c^\pdag_{Ln}-H.c.\right) =
i\left(\zeta_{1n}\xi_{1n}+\zeta_{2n}\xi_{2n}\right), 
\label{B:dimerization}
\end{equation}
which provides a relation between the dimerization operator and the 
Majorana's fermions.

\end{itemize}

\newpage
\section{Mean-field treatment of the $(\s, \tau)$-model}
\label{meanfield}

In oder to devise an appropriate mean-field scheme, 
let us formally rewrite Hamiltonian (\ref{H-new-rep}) as
\be
H[\s,\tau]=H_{mf}[\s] + H_{mf}[\tau] +
H_{fluc}[\s,\tau]+{\rm const.}
\label{staubis}
\ee
Here the two terms of the mean-field Hamiltonian are:
\bea
H_{mf}[\s] &=& - \sum_n \left( J_{\s} \s^z _n \s^z _{n+1}
+ \Delta_{\s} \s^x _n \right)
\label{mfa-sigma}\\
H_{mf}[\tau] &=&- \sum_n \left(J_{\tau} \tau^z _n \tau^z _{n+1}
+ h \tau^z _n + \Delta_{\tau}\tau^x _n  \right)
\label{mfa-tau}
\eea
where
\bea
J_{\s} = J \left(1 +  \la \tau^z _n \tau^z _{n+1} \ra \right), &&
J_{\tau} = J \la \s^z _n \s^z _{n+1} \ra - K \nonumber\\
\Delta_{\s} = \Delta \la \tau^x _n \ra - K, &&
\Delta_{\tau} = \Delta \left(1 + \la \s^x _n \ra \right)
\label{parametr.K=0}
\eea
The coupling between the fluctuations,
is given by
\bea
H_{fluc}[\s,\tau]&=&-\sum\limits_n
\left[J
(\s^z_{n}\s^z_{n+1}-\la\s^z_{n}\s^z_{n+1}\ra)
(\tau^z_{n}\tau^z_{n+1}-\la\tau^z_{n}\tau^z_{n+1}\ra)
\right.\nonumber\\&+&\left.
\Delta(\s^x_{n}-\la\s^x_n\ra)(\tau^x_{n}-\la\tau^x_n\ra)
\right]
\label{staufluc}
\eea

In the leading approximation, fluctuations described by (\ref{staufluc})
are neglected, and the $\s$ and $\tau$ degrees of freedom decouple.
We notice that the parameter $h$ appears only in $H_{mf}[\tau]$ where
it plays the role of an effective {\sl longitudinal} field. Therefore
the 
model (\ref{mfa-tau}) is a quantum counterpart of a 2D Ising model in
a nonzero external field which is known to be always noncritical. Thus,
as expected,
the $\tau$-chain is always gapped. 
The condition for the $\s$-chain to be critical then reads:
\be
J \left(1 +  \la \tau^z _n \tau^z _{n+1} \ra \right) =
\Delta \la \tau^x _n \ra - K
\label{crit-cond}
\ee

Despite the fact that the Ising model 
in the field is supposed to be integrable 
in the continuum limit (see references in \cite{MC}), 
the explicit expectation values
of the $\tau$ operators appearing in (\ref{crit-cond}) are not known.
To make further analytic progress possible, we choose
a special value of the field rescaling parameter
$K$, 
\be
K = J \la \s^z _n \s^z _{n+1} \ra, 
\label{special-K}
\ee
at which $J_{\tau} = 0$, and interaction between the neighbouring $\tau$-spins
in $H_{mf}[\tau]$ vanishes. This trivialises the $\tau$-model and leads to
\be
\la \tau^z \ra = \frac{h}{\sqrt{h^2 + \Delta^2 _{\tau}}},~~~
\la \tau^x \ra = \frac{\Delta _{\tau}}{\sqrt{h^2 + \Delta^2 _{\tau}}}
\label{tau-averages}
\ee
Self-consistency requires the knowledge of the averages related to the
$\s$-model. Specialising to the critical point and making use of the known
results \cite{Pf},
\be
\la \s^z _n \s^z _{n+1} \ra_{crit} = \la \s^x _n \ra_{crit} = \frac{2}{\pi},
\ee
we obtain
\be
K = \frac{2 J}{\pi}, ~~~\Delta_{\tau} = \Delta \left( 1 + \frac{2}{\pi}
\right).
\ee
Then the condition (\ref{crit-cond}) reduces to
\be
 1 + \la \tau^z \ra^2 = (\Delta / J)  \la \tau^x \ra -  \frac{2}{\pi} 
\label{eq-1.1}
\ee
This equation determines the critical value of $h$ at a given ratio
$\Delta / J$. 
Introducing the quantity 
$$
x = \frac{\sqrt{h^2 + C^2 \Delta^2}}{J}, ~~~C = 1 + \frac{2}{\pi}
$$
we obtain a quadratic equation
$$
(C+1) x^2 - C (\Delta/J)^2 x + C^2 (\Delta/J)^2 = 0,
$$
whose solution determines the critical value of $h$:
\be
\left(\frac{h_c}{C\Delta} \right)^2 
= \frac{1}{4 (C+1)^2} 
\left[\left(\frac{\Delta}{J}  \right)
+ \sqrt{\left(\frac{\Delta}{J}  \right)^2 + 4 (C+1)}\right]^2 - 1 
\label{crit-h}
\ee
The requirement that the r.h.side of (\ref{crit-h}) is positive yields a
restriction upon $\Delta$:
\be
\frac{\Delta}{J} > 1 + \frac{2}{\pi} \label{restr-ising} 
\ee
Recall that we already assumed that at $h = 0$ the two QI chains
of the QAT model are disordered: $\Delta > J$. 
The restriction (\ref{restr-ising})
tells us that, for the Ising criticality to be reached at some critical value
of $h$, the original disordered $\s_1$ and $\s_2$ chains should be 
far enough from their original critical point. In fact, as follows from
(\ref{restr-ising}), the value of the Majorana mass, when estimated as
$m \sim 2(\Delta - J) \sim 4J/\pi$, turns out to be of the order of the 
ultraviolet cutoff. This estimation reflects the already mentioned fact that 
the DM transition is
indeed not a weak-coupling one from the standpoint of the
DQAT model.

One could, in principle, perturbatively  calculate the
fluctuation corrections, i.e. those originating from (\ref{staufluc}),
to the criticality condition. 
As the model is  only tractable
under fine tuning (\ref{special-K}) of the parameter $K$,
that is not expected to generalise the qualitative picture obtained in the
mean-field approximation.

\newpage
\section{Correlation functions}
\label{corrf}

Let us start calculating the Fourier transform of $D^{(R)}(\omega,p)$,
given by (\ref{ret}),
with the remark that using the Bessel 
functions (despite some convenient Fourier transforms)
is {\sl beyond the accuracy}. The analytic properties
(and the issue of coherent particle poles) should not
depend on fine details of the correlators at 
$r\sim \xi=1/m$.
Therefore what one ought to do is to study analytic
properties of the integral
\be
I^m_\alpha(q)=\int\limits_0^\infty r^{\alpha-1/2} dr
J_0(qr)e^{-mr}
\label{int}
\ee
in terms of which the Fourier transforms,
\[
D(q)=\frac{1}{2}\int d^2\vec{r}
e^{i\vec{q}\vec{r}}D(r)=
\pi\int\limits_0^\infty r drJ_0(qr) D(r)\;,
\]
$\vec{q}=(\omega_n,p)$, $q=\sqrt{\omega_n^2+p^2}$, are
given by
\be
D_0(q)=\pi I^0_{5/4}(q),\;\;
D_>(q)=\frac{\pi^{3/2}A_1}{\sqrt{2m}}I^m_1(q),\;\;
D_<(q)=\frac{35A_1}{256 m^2}I^{2m}_{-1/2}(q)
\label{DI}
\ee

In the area of convergence, we have \cite{GR}:
\be
I^m_\alpha(q)=\frac{\Gamma(\alpha+1/2)}{m^{\alpha+1/2}}
F(\alpha/2+1/4,\alpha/2+3/4;1;-q^2/m^2)
\label{Igen}
\ee
where $F(a,b;c;z)$ stands for the hypergeometric
function. 
The character of the singularity at $z=1$ can be
determined by using  appropriate transformation
formulas for the hypergeometric functions 
and the $\Gamma$-function doubling formula\cite{GR}:
\be
I^m_\alpha(q)|_{q^2\simeq -m^2}=
\frac{(2m)^{\alpha-1/2}\Gamma(\alpha)}{\sqrt{\pi}}
\frac{1}{(q^2+m^2)^\alpha}
\label{Idiv}
\ee
where only the leading, most divergent term is
retained.

This calculation explains how the correlation function
has a branch cut at the threshold, unless $\alpha=1$
which, of course, corresponds to the disordered case:
\be
D_>(q)=\frac{\pi A_1}{q^2+m^2}
\label{Dorq}
\ee
where the coefficient, as expected, is the
same as one finds by Fourier transforming $K_0(mr)$
\[
K_0(mr)=\frac{1}{2\pi}\int d^2 \vec{q}
\frac{e^{i\vec{q}\vec{r}}}{q^2+m^2}
\]

Although this calculation is
instructive, formula (\ref{Idiv}) doesn't solve all
our problems.
So, the critical correlation function has nothing
to do with the $m\to 0$ limit of this formula,
as the main contribution to the integral 
comes from a different spatial domain. 
Therefore we should rather return to the general
formula
(\ref{Igen}), set $m=0$ there, and analytically 
continue beyond the convergence domain
($-1/2<\alpha<1/2$).
The result is
\be
D_0(q)= 2^{1/4}\pi \frac{\Gamma(7/8)}{\Gamma(1/8)}
\frac{1}{q^{7/4}}
\label{Dnotq}
\ee

The correlation function in the disordered phase
is also outside the validity of (\ref{Idiv}) and,
furthermore, the integral (\ref{Igen}) is
logarithmically
divergent at the lower limit in this case. It should
therefore be regularised. 
One way to do it is as follows \cite{GR}:
\bea
&~& I^{2m}_{-1/2}(q)=\int\limits_0^\infty \frac{dr}{r}
J_0(qr)e^{-2mr}\rightarrow \int_0^\infty \frac{dr}{r}
J_\nu(qr)
e^{-2mr} \nonumber\\
&=& \frac{(\sqrt{q^2+4m^2}-2m)^\nu}{\nu q^\nu}
\rightarrow \frac{1}{\nu}+\ln \left(
\frac{\sqrt{q^2+4m^2}-2m}{q}\right)+
{\rm O} (\nu)
\label{regp}
\eea
This regularisation is, however, not good enough. That
is because
the integral still diverges in the $q\to 0$ limit as
\[
\ln \left(\frac{q}{4m}\right)
\]
while the physical regularisation is {\it finite} 
in the $q\to 0$ limit and expands as:
\be
I^{2m}_{-1/2}(q)\rightarrow\int\limits_\xi^\infty
\frac{dr}{r}
J_0(qr)e^{-2mr}=\int\limits_\xi^\infty \frac{dr}{r}
e^{-2mr}+\sum\limits_{k=1}^{+\infty}\frac{(2k)!}{2^{2k}(k!)^2}
\left(-\frac{q^2}{4m^2}\right)^k
\label{regph}
\ee
Since different limiting processes must lead to the
same 
result, one can simply substruct the unphysical $q\to
0$ 
divergence from (\ref{regp}). This leads to the
following
result for the correlation function (up to an
unessential additive constant):
\be
D_<(q)=\frac{35A_1}{256m^2}\ln\left(\frac{4m}{\sqrt{q^2+4m^2}+2m}
\right)
\label{Ddisq}
\ee

Turning to the problem of analytic continuation, we recall that,
according to the standard approach \cite{AGD}
\[
D(i\omega_n,p)=D^{(R)}(\omega,p)
\]
in the upper half-plane. Put another way, one has to
substitute
\[
i\omega_n \to \omega+i\delta
\] 
(that is provided the resulting function has no
`accidental' poles and is otherwise regular in the
upper
half-plane - there is no general solution to the 
analytic continuation problem.)
The resulting function is, by construction, analytic
in the upper half-plane and automatically satisfies
the Kramers--Kronig relation.

We need to analytically continue formula
(\ref{Idiv}), or equivalently the function
\[
f(i\omega_n)=\frac{1}{(\omega_n^2+p^2+m^2)^\alpha}
\]
It is easy to see that
\bea
f(\omega)&=&
\frac{\theta(\epsilon_p^2-\omega^2)+\cos(\pi\alpha)
\theta(\omega^2-\epsilon_p^2)}{|\omega^2-\epsilon_p^2|^\alpha}
\nonumber\\ &-&
\frac{2i\sin(\pi\alpha)\theta(\omega^2-
\epsilon_p^2)}{|\omega^2-\epsilon_p^2|^\alpha}
\label{fan}
\eea


\end {document}